\documentclass[10pt,journal]{IEEEtran}
\ifCLASSOPTIONcompsoc
  \usepackage[nocompress]{cite}
\else
  \usepackage{cite}
\fi
\ifCLASSINFOpdf
\usepackage[pdftex]{graphicx}
\else
\usepackage[dvips]{graphicx}
\fi
\usepackage{epstopdf}
\usepackage{setspace}
\usepackage{diagbox}
\usepackage{multirow}
\usepackage{bbm}
\usepackage{cite}
\usepackage{graphicx,balance}
\usepackage{subcaption}
\usepackage{tabularx}
\usepackage{booktabs}
\usepackage{url}

\usepackage{amsthm}
\usepackage{amsmath}
\usepackage{amssymb}
\usepackage{dblfloatfix}
\usepackage{cases}
\usepackage{color}
\usepackage{amsfonts}

\makeatletter

\newcommand{\Rmnum}[1]{\expandafter\@slowromancap\romannumeral #1@}
\makeatother
\usepackage{algorithmic}
\usepackage{algorithm}
\usepackage{dsfont}
\usepackage{amssymb}
\usepackage{slashbox}
\allowdisplaybreaks[4]
\begin{document}
\title{\Huge {Age-Stratified COVID-19 Spread Analysis and Vaccination: A Multitype Random Network Approach}}

\author{Xianhao Chen,~\IEEEmembership{Student~Member,~IEEE,}
        Guangyu Zhu,~\IEEEmembership{Student~Member,~IEEE,}
        Lan Zhang,~\IEEEmembership{Member,~IEEE,}
        Yuguang Fang,~\IEEEmembership{Fellow,~IEEE},
        Linke Guo,  ~\IEEEmembership{Senior Member,~IEEE},
        and~Xinguang Chen
\thanks{Xianhao Chen, Guangyu Zhu, and Yuguang Fang are with the Department
of Electrical and Computer Engineering, University of Florida, Gainesville, FL 32611,
USA. (e-mail: xianhaochen@ufl.edu, gzhu@ufl.edu, fang@ece.ufl.edu).}
\thanks{Lan Zhang is with the Department of Electrical and Computer Engineering, Michigan Technological University, Houghton, MI 49931, USA, (e-mail: lanzhang@mtu.edu).}
\thanks{Linke Guo is with the Department
of Electrical and Computer Engineering, Clemson University, Clemson, SC 29634,
USA. (e-mail: linkeg@clemson.edu).}\thanks{Xinguang Chen is with the Department of Epidemiology, University of Florida, Gainesville, FL 32603, USA. (jimax.chen@ufl.edu).
}
\thanks{This work was supported in part by US National Science Foundation under grants IIS-1722791 and IIS-1722731. }}
\maketitle
\begin{abstract}
The risk for severe illness and mortality from COVID-19 significantly increases with age. As a result, age-stratified modeling for COVID-19 dynamics is the key to study how to reduce hospitalizations and mortality from COVID-19. By taking advantage of network theory, we develop an age-stratified epidemic model for COVID-19 in complex contact networks. Specifically, we present an extension of standard SEIR (susceptible-exposed-infectious-removed) compartmental model, called age-stratified SEAHIR (susceptible-exposed-asymptomatic-hospitalized-infectious-removed) model, to capture the spread of COVID-19 over multitype random networks with general degree distributions. We derive several key epidemiological metrics and then propose an age-stratified vaccination strategy to decrease the mortality and hospitalizations. Through extensive study, we discover that the outcome of vaccination prioritization depends on the reproduction number $R_0$. Specifically, the elderly should be prioritized only when $R_0$ is relatively high. If ongoing intervention policies, such as universal masking, could suppress $R_0$ at a relatively low level, prioritizing the high-transmission age group (i.e., adults aged 20-39) is most effective to reduce both mortality and hospitalizations. These conclusions provide useful recommendations for age-based vaccination prioritization for COVID-19.
\end{abstract}


\section{Introduction\label{sect: intro}}
Between January 2020 and November 30, 2020, about 1.47 million deaths from the novel coronavirus disease (COVID-19) are reported worldwide\cite{WHO}. On the one hand, COVID-19 is much more deadly than most strains of flu. On the other hand, many people infected with the coronavirus do not develop symptoms, and hence they can transmit the virus to others without being aware of it\cite{CDC}, which makes the pandemic extremely difficult to contain.

To live with the COVID-19 pandemic, governments and healthcare systems are always struggling to save lives and ``flatten the curve'', i.e., reducing the mortality and the peak of hospitalizations. Since severity and mortality rates of COVID-19 greatly vary across age-groups and increase dramatically for the elderly\cite{ferguson2020report,verity2020estimates}, effective intervention policies to achieve these two goals must prevent elderly, who are at high-risk for severe clinical outcomes, from infections. For this reason, age-stratified modeling for COVID-19 dynamics indeed serves as the basis of accurately assessing the effectiveness of control policies in decreasing illness severity and mortality. In this respect, some age-stratified mathematical models have already been proposed to analyze the spread of COVID-19 for different purposes\cite{singh2020age,balabdaoui2020age,tuite2020mathematical,jentsch2020prioritising}. However, these models are based on an oversimplified assumption that people are fully mixing, i.e., everyone contracting and spreading the virus to every other with equal probability, within each age group, which clearly fail to incorporate enough details in real-life contact networks. In reality, people in the same age group still differ greatly in the way of spreading the disease. As a consequence of this heterogeneity, it is found that epidemic outcomes in complex networks could deviate greatly from the results obtained from fully mixing epidemiological models\cite{meyers2005network,hebert2020beyond}.

Motivated by the aforementioned observations, in this paper, we present a unified yet simple mathematical model for COVID-19 spread analysis by accounting for both the age-specific risk and the heterogeneity in contact patterns within and across age groups. We take advantage of random network theory to analyze the spread of COVID-19 in contact networks with general degree distributions. More specifically, we present an extension of standard SEIR (susceptible-exposed-infectious-removed) compartmental model, called age-stratified SEAHIR (susceptible-exposed-asymptomatic-hospitalized-infectious-removed) model to describe the disease progression for infected individuals, and study the epidemic spreading process in multitype random networks where each type of nodes is treated as an age group. Some key epidemiological metrics, such as time-dependent dynamics, steady-state epidemic size (which will be termed as epidemic size throughout this paper), epidemic probability, and reproduction number, are derived, allowing us to analyze the epidemics and the impact of control policies in a thorough and effective manner. Due to the consideration of stochasticity and network structure, the proposed model is capable of offering some useful epidemic results that the existing fully mixing age-stratified models are unable to provide, like assessing the impact of preferential isolation of nodes (e.g., immunizing essential workers first). Given that many contagious diseases, including influenza, also exhibit distinct characteristics for different groups of people\cite{medlock2009optimizing}, the proposed model can be easily generalized to modeling many other infectious diseases.

\begin{figure}[t]
\centering
\includegraphics[width=2.5in]{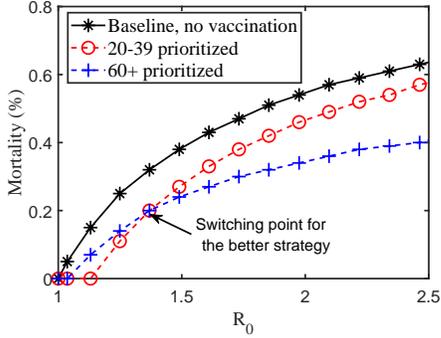}
\caption{Mortality (the ultimate death toll over the whole population) versus reproduction number $R_0$ with $10\%$ vaccination coverage. Here, $R_0$ denotes the reproduction number prior to vaccination, where we control the level of NPIs to vary $R_0$ (the details are given in Section \ref{subsect: settings}). ``20-39 prioritized'' or ``60+ prioritized'' means that these vaccine doses (enough to vaccinate 10\% population) are uniformly given to the adults aged 20-39 or aged 60+. The simulation is based on real-world age-stratified contact matrix for the United States\cite{prem2017projecting}.}
\label{fig: vaccineR0}
\end{figure}

While non-pharmaceutical intervention (NPI) policies, such as masking and social distancing, are effective in reducing the transmissions and mitigating the healthcare burden, it has become increasingly clear that vaccination is the only way to eliminate the pandemic worldwide. Unfortunately, vaccine availability will be highly constrained for general population during at least the first several months of the vaccine distribution campaign. Therefore, vaccination prioritization decision will play a pivotal role in reducing the effects of COVID-19 during such a period\cite{CDCvaccine,vaccinechallenges}. Under our proposed framework, we present an age-stratified vaccination strategy for the considered multitype network. In simulations, we focus on answering the following question: with limited doses available, who should be vaccinated first to reduce mortality and hospitalizations as much as possible? Our simulation results show that the answer depends on the value of reproduction number $R_0$. The reason behind is that, the epidemic size (i.e., the fraction of population eventually getting infected) increases slowly in large $R_0$ region, while increasing steeply in small $R_0$ region. As a result, vaccinating the high-transmission group (adults aged 20-39) is highly effective in blocking COVID-19 transmissions in small $R_0$ region, which thus protects the high-risk group (the elderly) indirectly. In contrast, in large $R_0$ region, even if high-transmission group is prioritized, it will have little impact on epidemic size as long as the vaccine supply is limited. Consequently, directly vaccinating the high-risk group becomes the preferable strategy. We illustrate this phenomenon in Fig. \ref{fig: vaccineR0}, where prioritizing young people aged 20-39 is preferable when $R_0<1.36$, whereas prioritizing the elderly is the better choice only when $R_0>1.36$. Although most studies estimate that $R_0$ for COVID-19 is between 2-3.5 under pre-intervention scenarios\cite{britton2020mathematical}, it certainly can be pushed to a relatively low level, e.g., below $1.36$, via NPI policies or even natural immunity (the latter only meaningful for highly infected places\cite{tkachenko2020persistent}). Thus, our finding indicates that vaccination prioritization should be customized for different places by considering the ongoing NPI policies and other effects that could suppress $R_0$. The key contributions of this paper are summarized as follows.

\begin{itemize}
\item We employ multitype random network theory to develop an age-stratified epidemic model for COVID-19. We derive the time-dependent epidemic dynamics, where each individual could belong to one of six compartments, i.e., susceptible, exposed, asymptomatic, hospitalized, infectious and removed.

\item To analyze the stochastic property and final state of the epidemic, we derive other critical epidemiological metrics, such as epidemic size, epidemic probability, and reproduction number for the considered networks.

\item We present an age-stratified vaccination strategy based on the proposed model. The simulation results indicate that high-risk age group should be vaccinated first to diminish mortality and hospitalizations in large $R_0$ region. Conversely, when $R_0$ is suppressed at a low level, prioritizing the high-transmission age group becomes the most effective strategy.
\end{itemize}

The reminder of this paper is organized as follows. In Section \ref{sect: Related}, we describe the related work. In Section \ref{sect: analysis}, we introduce the network model, and derive the time-dependent epidemic dynamics and other key epidemiological metrics. In Section \ref{sect: vaccination}, we devise an vaccination strategy for the considered networks. In Section \ref{sect: simulation}, we conduct simulations to compare different age-specific vaccination prioritization strategies. In Section \ref{sect: conclusion}, we draw our conclusions.

\section{Related Work\label{sect: Related}}
Some mathematical models for COVID-19 have been presented to account for the age-varying risks for mortality and severe illness. In \cite{singh2020age}, Singh et al. use an age-stratified SIR (susceptible-infective-removed) model to study the impact of social distancing measures, including workplace non-attendance, school closure, and lockdown, on the course of the COVID-19 pandemic. In \cite{balabdaoui2020age}, Balabdaoui et al. propose an age-stratified discrete compartmental model to describe the day-by-day progression of an infected individual in modern healthcare systems, e.g., in intensive care unit (ICU), with the objective of precisely projecting the occupancy of health care resources. In \cite{tuite2020mathematical}, Tuite et al. develop an age-stratified COVID-19 model to identify intervention strategies that keep the number of projected severe cases lower than the capacity of local health care systems. The aforementioned models make full-mixing assumption within each age group, which hence fail to capture enough details of population heterogeneity. In \cite{chang2020modelling}, Chang et al. propose an agent-based model to predict the infected number in Australia by considering the age-dependent effects. While agent-based models incorporate more realistic factors, they demand computationally intensive simulations, and generally offer limited insights into epidemic outcomes.

Random network theory allows us to model epidemics by taking heterogeneous contact network structure into account while bypassing computationally complicated simulations. Epidemic propagation in networks can be exactly interpreted as a bond percolation process, which hence can be analyzed by well-understood physics models, such as percolation\cite{newman2002spread}. Although several works have applied percolation theory to analyze the spread of COVID-19\cite{hebert2020beyond,allard2020role,chen2020time}, they have not taken the age-varying effects into consideration. On the other hand, given that an age-stratified population can be characterized as a multitype random network in which each type of vertices correspond to an age group, one possible direction is to directly map the epidemic spread to bond percolation in multitype random graphs\cite{allard2009heterogeneous,allard2015general}. Unfortunately, percolation theory is mostly limited to analysis of final state of networks, and cannot predict time-dependent transient dynamics. In \cite{miller2012edge}, Miller et al. propose an edge-based SIR compartmental model to describe the time-dependent epidemic dynamics in complex networks. Inspired by their approach, we solve the time-dependent dynamics for COVID-19 in multitype random networks, and then derive the expressions for epidemic size, epidemic probability, and reproduction number by performing analysis on the final state of the considered networks.

The design of vaccination prioritization strategies for COVID-19 has also attracted some research attention. Nonetheless, most of works draw the conclusion that vaccinating the older groups first is the robust strategy to minimize mortality or hospitalizations during a vaccine shortage\cite{matrajt2020vaccine,bubar2020model,buckner2020optimal}. This perhaps is because they fail to identify the underlying relationship between the priority population and the reproduction number $R_0$. Our finding coincides with these works only when $R_0$ is great. Recently, Jentsch et al. show that prioritizing the high-transmission group will reduce the death toll from COVID-19 most if vaccines become available late next year for Ontario, because high level of natural immunity may be already achieved in Ontario at that time\cite{jentsch2020prioritising}. Their conclusion essentially shares the same observation with ours as higher natural immunity leads to a lower $R_0$. Different from their work, by taking advantage of our epidemic model, we also study the impact of vaccination prioritization strategies on hospitalizations, and the effectiveness of immunizing people with high activity, i.e., the essential workers. Furthermore, our simulation results show that vaccinating high-transmission group is highly effective as long as $R_0$ is small, which applies to areas that are either hit hard as in \cite{jentsch2020prioritising} or only have few infections but with relatively strict NPI policies, e.g., masking mandate.

\section{Epidemic analysis\label{sect: analysis}}
\subsection{Network and compartmental model}
Let us consider a \textit{multitype} network which consists of $M$ types of nodes, \textit{each corresponding to an age group in a population}. We use $w_i$ to represent the fraction of the nodes of type $i\in[1, M]$. The contact from a type-$i$ node to others follows degree distribution $p_i(k_1,k_2,...,k_M)\triangleq p_i(\boldsymbol k)$, describing the joint probability for type-$i$ node to be connected with $k_1$ type-$1$ node, $k_2$ type-$2$ node, ..., and $k_M$ type-$M$ node, where ${\boldsymbol k} =(k_1, k_2, \ldots, k_M)$. The considered network can be generated by the following procedure: 1) generate stubs for every node following degree distribution $p_i(\boldsymbol k)$, where each stub contains the information about which type of node it reaches. 2) randomly wire two matching stubs together to create an edge and repeat this process until no stubs left.

Susceptible-Infected-Removed (SIR) and Susceptible-Exposed-Infected-Removed (SEIR) compartmental models are widely used for epidemic modeling. In SEIR compartmental model, each individual can be in one of the four states, i.e., Susceptible, Exposed, Infected, or Removed. Here, to capture the salient features of COVID-19, we present a novel compartmental model, i.e., SEAHIR model, which adds two additional compartments, i.e., asymptomatic and hospitalized, to the classic SEIR model. In SEAHIR model, each individual can be in one of the six states: susceptible (S), exposed (E), symptomatic and infectious (I), asymptomatic and infectious (A), hospitalized (H), and removed (R). Both new compartments are paramount to describe the dynamics of COVID-19: the number of people in H state indicates the hospitalizations, which must be kept lower than health care capacity; patients in A state have different level of infectivity compared with symptomatic ones\cite{li2020substantial}. We assume that individuals in E state is not infectious because of low virus load, and individuals in H state are properly isolated. Individuals in I and A states are assumed to be infectious to others, where the infection rate from a type-$i$ source node to a type-$j$ node is $\lambda_{i,j}^I$ or $\lambda_{i,j}^A$ given the source node belongs to I or A state. For conciseness, we do not distinguish recovery and death in R state, but assume that an age-dependent fraction of infected people will die. Furthermore, given the fact that the cases of reinfection with COVID-19 are still extremely rare, we do not consider the transition from R state to S state. For type-$i$ nodes, the transitions among the compartments are illustrated in Fig. \ref{fig: SEAHIR}, where the symbols on the arrows denote the corresponding transition rates from one to another, which are all dependent on node type $i$ to account for the age-dependent effects.

\subsection{Time-dependent dynamics\label{sect: dynamics}}
\begin{figure}[t]
\centering
\includegraphics[width=2.7in]{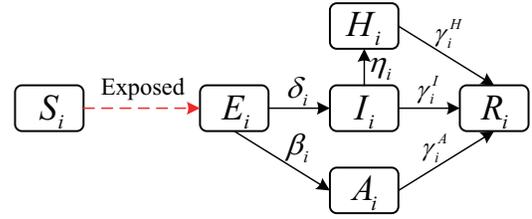}
\caption{SEAHIR compartmental model for nodes of type $i$.}
\label{fig: SEAHIR}
\end{figure}

Disease propagates from infectious nodes to its neighbors, leading to an epidemic if the epidemic size is comparable to the whole population. We employ the edge-based compartmental method to solve the equations of dynamics governing the epidemic spread over random graph\cite{miller2012edge}. The core idea of the edge-based method is to shift our attention from an individual node to \textit{the status of its neighbor reached by an edge}. To study the impact of vaccination or natural immunity, we consider that a fraction of population may be already immune at the beginning of analysis: $S_i(\boldsymbol k,0)$ represents the fraction of type-$i$ nodes with degree $\boldsymbol k$ that are initially susceptible. Besides, we use $\theta_{ji}(t)$ to denote the probability that a type-$j$ neighbor has \textit{not} transmitted the disease to a type-$i$ node by time $t$ given that the type-$i$ node is susceptible at time $0$. $\theta_{ji}(t)$ can be interpreted as a state of a type-$j$ neighbor of an initially susceptible type-$i$ node. It is noted that $\theta_{ji}(0)=1$ according to the definition. Based on Fig. \ref{fig: SEAHIR}, we can construct the following equations to characterize the time-dependent epidemic process.
\begin{align}
S_i(t)&=\sum_{\boldsymbol k} S_i(\boldsymbol k,0) p_i(\boldsymbol k) \prod_{l=1}^{M}\theta^{k_l}_{li}(t),\label{d1}\\
\dot A_i(t)&=\beta_i E_i(t) - \gamma_i^A A_i(t),\label{d2}\\
\dot I_i(t)&=\delta_i E_i(t) - \eta_i I_i(t) - \gamma_i^I I_i(t),\label{d3}\\
\dot H_i(t)&=\eta_i I_i(t) - \gamma_i^H H_i(t),\label{d4}\\
\dot R_i(t)&=\gamma_i^A A_i(t) + \gamma_i^I I_i(t) + \gamma_i^H H_i(t),\label{d5}\\
E_i(t)&=1 - S_i(t) - A_i(t) - I_i(t) - H_i(t) - R_i(t),\label{d6}
\end{align}
where $S_i(t)$, $A_i(t)$, $I_i(t)$, $H_i(t)$, $R_i(t)$, and $E_i(t)$ represent the proportions of type-$i$ nodes in the corresponding states at time $t$, respectively.  From Markov chain theory, when the network size is sufficiently large, the fraction of nodes in $A_i$, $I_i$, $H_i$, and $R_i$ states can be described well by the differential equations (\ref{d2})-(\ref{d5}) due to the flow diagram in Fig. \ref{fig: SEAHIR}. Moreover, $(\ref{d6})$ is obtained from $S_i(t)+A_i(t)+I_i(t)+H_i(t)+R_i(t)+E_i(t)=1$. For the initial conditions, we assume $A_i(0)=I_i(0)=H_i(0)=E_i(0)=0$ and $R_i(0)=1-S_i(0)$. Obviously, one can solve the above equations as long as the key probability, i.e., $\theta_{ji}(t)$, is derived. To calculate $\theta_{ji}(t)$, following the approach in \cite{miller2012edge}, we break it into six parts, i.e., $\xi_{ji}^S(t)$, $\xi_{ji}^E(t)$, $\xi_{ji}^A(t)$, $\xi_{ji}^I(t)$, $\xi_{ji}^H(t)$, and $\xi_{ji}^R(t)$. Specifically, $\xi_{ji}^S(t)$, $\xi_{ji}^E(t)$, $\xi_{ji}^A(t)$, $\xi_{ji}^I(t)$, $\xi_{ji}^H(t)$, or $\xi_{ji}^R(t)$ represents the probability that the considered type-$j$ neighbor is in $S$, $E$, $A$, $I$, $H$, or $R$ state, respectively, and has \textit{not} transmitted the disease to the initially susceptible type-$i$ node by time $t$, satisfying
\begin{gather}
\theta_{ji}(t)=\xi_{ji}^S(t)+\xi_{ji}^E(t)+\xi_{ji}^A(t)+\xi_{ji}^H(t)+\xi_{ji}^I(t)+\xi_{ji}^R(t). \label{theta}
\end{gather}

A type-$j$ neighbor in $S$ state cannot infect the type-$i$ node, and hence $\xi_{ji}^S(t)$ is simply equal to the probability that the considered type-$j$ neighbor is susceptible. The degree distribution of the considered type-$j$ neighbor is given by $\frac{k_i p_j(\boldsymbol k)}{\overline k_{ji}}$, where $\overline k_{ji} = \sum_{\boldsymbol k} k_i p_j(\boldsymbol k)$ is the average degree leaving from type-$j$ node to type-$i$ node, which normalizes the probability distribution. This quantity is proportional to $k_i p_j(\boldsymbol k)$ because type-$j$ nodes with more edges incident to type-$i$ nodes are more likely to become the neighbors of type-$i$ nodes\cite{allard2009heterogeneous}. We can obtain $\xi_{ji}^S(t)$ by computing the probability that the type-$j$ neighbor is initially susceptible and has not been infected by any of its neighbors, except the considered type-$i$ node, by time $t$, i.e.,
\begin{gather}
\xi_{ji}^S(t)=\sum_{\boldsymbol k} \frac{k_i S_j(\boldsymbol k,0) p_j(\boldsymbol k)}{\overline k_{ji}} \prod_{l=1}^{M}\theta^{k_l-\delta_{il}}_{lj}(t),\label{S1}
\end{gather}
where $\delta_{il}$ is the Kronecker delta operator, with $\delta_{il}=1$ only when $i=l$, and $\delta_{il}=0$ otherwise.

As mentioned before, two states in our compartmental model, i.e., $I$ and $A$ states, are infectious. Thus, the decrease in $\theta_{ji}$ comes from two joint events: 1) the type-$j$ neighbor is $A$ or $I$ state, and 2) it transmits the disease to the type-$i$ node with infection rate $\lambda_{ji}^A$ or $\lambda_{ji}^I$, i.e.,
\begin{gather}
-\dot \theta_{ji}(t)= \big(\lambda_{ji}^A\xi_{ji}^A(t) + \lambda_{ji}^I\xi_{ji}^I(t)\big).\label{thetaderivative}
\end{gather}

One can also interpret (\ref{thetaderivative}) in this way: there is a state $1-\theta_{ji}(t)$ (which corresponds to that the type-$j$ neighbor \textit{has} transmitted the disease to the initially susceptible type-$i$ node by time $t$) receiving the flows from both $\xi_{ji}^A(t)$ and $\xi_{ji}^I(t)$.

If staying in $A$, $I$ or $H$ state, the type-$j$ neighbor transits to R state with rate $\gamma_i^A$, $\gamma_i^I$ or $\gamma_i^H$, which means
\begin{gather}
\dot \xi_{ji}^R(t)=\gamma_j^A \xi_{ji}^A(t) + \gamma_j^I \xi_{ji}^I(t) + \gamma_j^H \xi_{ji}^H(t).
\end{gather}

The type-$j$ neighbor progresses from I state to H state with rate $\eta_j$, leading to
\begin{gather}
\dot \xi_{ji}^H(t)=\eta_j \xi_{ji}^I(t) - \gamma_j^H \xi_{ji}^H(t).\label{xih}
\end{gather}

States $\xi_{ji}^A(t)$ and $\xi_{ji}^I(t)$ receive the flows from state $\xi_{ji}^E(t)$. Besides, $\xi_{ji}^A(t)$ progresses to state $1-\theta_{ji}(t)$ and $\xi_{ji}^R(t)$, while $\xi_{ji}^I(t)$ progresses to state $1-\theta_{ji}(t)$, $\xi_{ji}^R(t)$, and $\xi_{ji}^H(t)$, yielding
\begin{align}
\dot \xi_{ji}^A(t)&= \beta_j\xi_{ji}^E(t) - \gamma_j^A \xi_{ji}^A(t) -\lambda_{ji}^A\xi_{ji}^A(t),\\
\dot \xi_{ji}^I(t)&= \delta_j\xi_{ji}^E(t) - \gamma_j^I \xi_{ji}^I(t) - \eta_j \xi_{ji}^I(t) - \lambda_{ji}^I\xi_{ji}^I(t).
\end{align}

In summary, one can solve $\theta_{ji}(t)$ from the following equations.

\begin{numcases}{}
\begin{split}
\xi_{ji}^S(t)&=\sum_{\boldsymbol k} \frac{k_i S_j(\boldsymbol k,0) p_j(\boldsymbol k)}{\overline k_{ji}} \prod_{l=1}^{M}\theta^{k_l-\delta_{il}}_{lj}(t), \\
\dot \theta_{ji}(t)&= - \big(\lambda_{ji}^A\xi_{ji}^A(t) + \lambda_{ji}^I\xi_{ji}^I(t)\big),\\
\dot \xi_{ji}^A(t)&= \beta_j\xi_{ji}^E(t) - \gamma_j^A \xi_{ji}^A(t) -\lambda_{ji}^A\xi_{ji}^A(t),\\
\dot \xi_{ji}^I(t)&= \delta_j\xi_{ji}^E(t) - \gamma_j^I \xi_{ji}^I(t) - \eta_j \xi_{ji}^I(t)-\lambda_{ji}^I\xi_{ji}^I(t),\\
\dot \xi_{ji}^H(t)&=\eta_j \xi_{ji}^I(t) - \gamma_j^H \xi_{ji}^H(t),\\
\dot \xi_{ji}^R(t)&=\gamma_j^A \xi_{ji}^A(t) + \gamma_j^I \xi_{ji}^I(t)+ \gamma_j^H \xi_{ji}^H(t),\\
\xi_{ji}^E(t)&=\theta_{ji}(t)- \xi_{ji}^S(t)-\xi_{ji}^A(t)-\xi_{ji}^H(t)-\xi_{ji}^I(t)\\
&-\xi_{ji}^R(t),
\end{split}
\label{thetaall}
\end{numcases}
where $\xi_{ji}^R(0) =  1 - \sum_{\boldsymbol k} \frac{k_i S_j(\boldsymbol k,0) p_j(\boldsymbol k)}{\overline k_{ji}}$, $\theta_{ji}(0)=1$, and $\xi_{ji}^A(0)=\xi_{ji}^I(0)=\xi_{ji}^H(0)=0$. By plugging $\theta_{ji}(t)$ into (\ref{d1}), we can obtain the fraction of type-$i$ nodes in each compartment at given time from (\ref{d1})-(\ref{d6}). As a result, the desired age-stratified epidemic dynamics can be obtained by solving $\mathcal O(M^2)$ equations, where $M$ is the number of age groups. One can see that (\ref{d1})-(\ref{d6}) and (\ref{thetaall}) account for considerably more population structure than a fully mixing model by only introducing marginally more complexity.
\subsection{Epidemic size and epidemic probability}
Epidemic size measures the fraction of people eventually getting infected, and epidemic probability is defined as the likelihood that the first infected patient sparks an epidemic\cite{meyers2005network}. SEAHIR model contains more compartments than traditional SIR or SEIR models, which complicates the derivations of these two key metrics. Fortunately, both metrics only depend on final state of networks. From Fig. \ref{fig: SEAHIR}, it is intuitive to see that the network progresses to an equilibrium when $t\rightarrow\infty$, where $E_i(\infty)=A_i(\infty)=I_i(\infty)=H_i(\infty)=0$. By using this fact, we use a single compartment $\mathbb I$ to replace all the infected states, i.e., E, A, I, H states in the original SEAHIR compartmental model. Here is our trick: although S$\mathbb I$R model cannot be used to capture the temporal dynamics of SEAHIR model, it can be calibrated appropriately to share the same final network state, i.e., $S_i(\infty)$ and $R_i(\infty)$, with SEAHIR model. Let $T_{ji}\in[0, 1]$ denote the probability that an infected type-$j$ node ultimately transmits the disease to an initially susceptible adjacent type-$i$ node. We assume that the S$\mathbb I$R model is with infection rate $\hat\lambda_{ji}$ from type-$j$ node to type-$i$ node and transition rate $\hat \gamma_{ji}$ from $\mathbb I$ state to R state for a type-$j$ neighbor of a type-$i$ node. Since $T_{ji}$ for the S$\mathbb I$R model is $\frac{\hat\lambda_{ji}}{\hat\lambda_{ji}+\hat\gamma_{ji}}$, the S$\mathbb I$R model has exactly the same final network state as the SEAHIR model as long as $\frac{\hat\lambda_{ji}}{\hat\lambda_{ji}+\hat\gamma_{ji}}$ is set to the $T_{ji}$ in the SEAHIR model.

Let us first calculate $T_{ji}$ in the SEAHIR model, and then derive the desired metrics based on the S$\mathbb I$R model. An infected type-$j$ node may enter one of two infectious states, i.e., $I$ state or $A$ state, with probability $\frac{\delta_j}{\beta_j+\delta_j}$ or $\frac{\beta_j}{\beta_j+\delta_j}$. Given that the type-$j$ node in I or A state, suppose that there are two stages that it will progress to, i.e., ``infecting the adjacent type-$i$ node'' and ``leaving current state'', with rates $\lambda_{ji}^I$ (or $\lambda_{ji}^A$) and $\eta_j+\gamma^I_j$ (or $\gamma^A_j$), respectively. $T_{ji}$ is the probability that the considered node enters the first stage, which equals $\frac{\lambda^I_{ji}}{\lambda^I_{ji}+(\eta_j+\gamma^I_j)}$ if the type-$j$ node is in $I$ state, and equals $\frac{\lambda^A_{ji}}{\lambda^A_{ji}+\gamma^A_j}$ if it is in $A$ state, yielding
\begin{align}
T_{ji}=&\frac{\delta_j}{\beta_j+\delta_j} \frac{\lambda^I_{ji}}{\lambda^I_{ji}+(\eta_j+\gamma^I_j)}+ \frac{\beta_j}{\beta_j+\delta_j}\frac{\lambda^A_{ji}}{\lambda^A_{ji}+\gamma^A_j}. \label{Tij}
\end{align}

In S$\mathbb I$R model, we use $\hat \xi_{ji}^S(t)$, $\hat \xi_{ji}^{\mathbb I}(t)$, or $\hat \xi_{ji}^R(t)$ to represent the probability that a type-$j$ neighbor of an initially susceptible type-$i$ node is in $S$, $\mathbb I$, or $R$ state and meanwhile has not infected this type-$i$ node by time $t$, and use $\hat \theta_{ji}(t)$ to denote the probability that the type-$j$ neighbor has not infected the initially susceptible type-$i$ node by time $t$, satisfying $\hat \theta_{ji}(t)=\hat\xi_{ji}^S(t)+\hat\xi_{ji}^{\mathbb I}(t)+\hat\xi_{ji}^R(t)$. Notice that $\hat \theta_{ji}(t)=\theta_{ji}(t)$ only when $t=\infty$, because SEAHIR and S$\mathbb I$R model share the same final network state while having different temporal dynamics. By analogy with the preceding subsection, we can obtain
\begin{align}
\hat\xi_{ji}^S(t)=&\sum_{\boldsymbol k} S_j(\boldsymbol k,0) \frac{k_i p_j(\boldsymbol k)}{\overline k_{ji}} \prod_{l=1}^{M}\hat\theta^{k_l-\delta_{il}}_{lj}(t),\label{S2}\\
\hat\xi_{ji}^R(t)=&\frac{(1-\hat\theta_{ji}(t))\hat\gamma_{ji}}{\hat\lambda_{ji}}+\hat\xi_{ji}^R(0),\label{R2}\\
\dot{\hat \theta}_{ji}(t)=&- \hat\lambda_{ji}\hat\xi_{ji}^{\mathbb I}(t),\label{theta2}\\
\hat\xi_{ji}^{\mathbb I}(t)=&\hat\theta_{ji}(t)-\hat\xi_{ji}^S(t)-\hat\xi_{ji}^R(t)\notag\\
=&\hat\theta_{ji}(t)-\sum_{\boldsymbol k} S_j(\boldsymbol k,0) \frac{k_i p_j(\boldsymbol k)}{\overline k_{ji}} \prod_{l=1}^{M}\hat\theta^{k_l-\delta_{il}}_{lj}(t)\notag\\
&-\frac{(1-\hat\theta_{ji}(t))\hat\gamma_{ji}}{\hat\lambda_{ji}}-\hat\xi_{ji}^R(0),\label{I2}
\end{align}
where $\hat\xi_{ji}^R(0) =  1 - \sum_{\boldsymbol k} \frac{k_i S_j(\boldsymbol k,0) p_j(\boldsymbol k)}{\overline k_{ji}}$ is the probability that the type-$j$ neighbor is initially removed (immune), and $\hat \theta_{ji}(0)=1$. The derivation of (\ref{S2}) and (\ref{theta2}) is similar to that of (\ref{S1}) and (\ref{thetaderivative}). Analogous to the preceding subsection, probability $1-\hat\theta_{ji}(t)$ corresponds to that the type-$j$ neighbor \textit{has} transmitted the disease to the initially susceptible type-$i$ node by time $t$. State $\hat \xi_{ji}^{\mathbb I}(t)$ transits to state $1-\hat\theta_{ji}(t)$ with rate $\hat\lambda_{ji}$ while transiting to state $\hat\xi_{ji}^R(t)$ with rate $\hat\gamma_{ji}$, leading to the relationship between $\hat\xi_{ji}^R(t)$ and $1-\hat\theta_{ji}(t)$ in (\ref{R2}).

Clearly, the epidemic dynamics can be governed by (\ref{theta2}) and (\ref{I2}). By taking (\ref{I2}) into (\ref{theta2}), we have
\begin{align}
\dot{\hat \theta}_{ji}(t)=& \hat\lambda_{ji} \hat\xi_{ji}^R(0) + \hat\lambda_{ji} \sum_{\boldsymbol k} S_j(\boldsymbol k,0) \frac{k_i p_j(\boldsymbol k)}{\overline k_{ji}} \prod_{l=1}^{M}\hat\theta^{k_l-\delta_{il}}_{lj}(t)\notag\\
&+(1-\hat\theta_{ji}(t))\hat\gamma_{ji}-\hat\lambda_{ji}\hat\theta_{ji}(t). \label{thetasystem}
\end{align}

Now the advantage of using S$\mathbb I$R compartmental model becomes clearer: we arrive at (\ref{thetasystem}) which is only related to $\hat\theta_{ji}(t)$. Since the population is closed, the epidemic will eventually go extinct, implying $\hat\xi_{ji}^{\mathbb I}(\infty)=0$. Given that $\dot{\hat \theta}_{ji}(\infty)=- \hat\lambda_{ji}\hat\xi_{ji}^{\mathbb I}(\infty)=0$ and $\theta_{ji}(\infty)=\hat\theta_{ji}(\infty)$, and recall that $\frac{\hat\lambda_{ji}}{\hat\lambda_{ji}+\hat\gamma_{ji}} = T_{ji}$, we can go back to the original SEAHIR model and obtain
\begin{align}
\theta_{ji}(\infty)=& T_{ji}\Big(1-\sum_{\boldsymbol k} S_j(\boldsymbol k,0) \frac{k_i p_j(\boldsymbol k)}{\overline k_{ji}} \big(1-\prod_{l=1}^{M}\theta^{k_l-\delta_{il}}_{lj}(\infty)\big)\Big)\notag\\
&+1-T_{ji},\label{self1}
\end{align}
where $T_{ji}$ can be calculated from (\ref{Tij}). One can solve $\theta_{ji}(\infty)$ from the above equation. In fact, (\ref{self1}) can be explained in an intuitive way: a type-$i$ node has not been infected by a type-$j$ neighbor since $t=0$ either because it cannot be reached from its neighbor with probability $1-T_{ji}$, or it can be reached from its neighbor with probability $T_{ji}$ but the neighbor has not been infected since $t=0$. Thus, as $t\rightarrow\infty$, the fraction of type-$i$ nodes that have been infected since $t=0$ is given by $R_i(\infty)-R_i(0)$, i.e.,
\begin{align}
&R_i(\infty)-R_i(0)=\big(1-\sum_{\boldsymbol k} S_i(\boldsymbol k,0) p_i(\boldsymbol k) \prod_{l=1}^{M}\theta^{k_l}_{li}(\infty)\big)-\notag\\
&\big(1-\sum_{\boldsymbol k} S_i(\boldsymbol k,0) p_i(\boldsymbol k)\big)=\sum_{\boldsymbol k} S_i(\boldsymbol k,0) p_i(\boldsymbol k) \big(1-\prod_{l=1}^{M}\theta^{k_l}_{li}(\infty)\big).
\end{align}

The epidemic size is therefore expressed as
\begin{align}
\mathcal R=&\sum_{i=1}^M w_i \Big(\sum_{\boldsymbol k} S_i(\boldsymbol k,0) p_i(\boldsymbol k) \big(1-\prod_{l=1}^{M}\theta^{k_l}_{li}(\infty)\big)\Big).\label{size}
\end{align}

It is noted that when the population is fully susceptible, i.e., $S_i(\boldsymbol k,0)=1$ for all $j$ and $\boldsymbol k$, $\mathcal R$ solved from (\ref{self1}) and (\ref{size}) is the size of giant component in multitype networks obtained by percolation theory\cite{allard2009heterogeneous,allard2015general}. We can compute epidemic probability in a similar way. Let $\mu_{ij}(\infty)$ denote the probability that a type-$i$ node (which is assumed to be infected) does not spark an epidemic via a type-$j$ neighbor. Analogous to (\ref{self1}), the following argument is true: a type-$i$ node does not ignite an epidemic via a type-$j$ node either because it cannot infect its type-$j$ neighbor, or it can infect its type-$j$ neighbor but the latter cannot spark an epidemic, which means
\begin{align}
\mu_{ij}(\infty)=&T_{ij}\Big(1- \sum_{\boldsymbol k} S_j(\boldsymbol k,0) \frac{k_i p_j(\boldsymbol k)}{\overline k_{ji}} \big(1-\prod_{l=1}^{M}\mu^{k_l-\delta_{il}}_{jl}(\infty)\big)\Big)\notag\\
&+1-T_{ij}. \label{self2}
\end{align}

As a result, the probability that an infected type-$i$ node ignites an epidemic is given by
\begin{gather}
\mathcal P_i=1 - \sum_{\boldsymbol k} p_i(\boldsymbol k) \prod_{l=1}^{M}\mu^{k_l}_{il}(\infty). \label{prob1}
\end{gather}

By randomly choosing a susceptible node to infect, we define the likelihood that it starts an epidemic as the epidemic probability:
\begin{align}
\mathcal P=&\sum_{i=1}^M \overline w_i \mathcal P_i, \label{prob2}
\end{align}
where $\overline w_i= \frac{\sum_{\boldsymbol k} w_i S_i(\boldsymbol k,0)}{\sum_i \sum_{\boldsymbol k} w_iS_i(\boldsymbol k,0) }$ is the probability that the randomly chosen susceptible node is of type $i$, which reduces to $w_i$ in a fully susceptible population.
\subsection{Reproduction number}
One of the fundamental parameters for an epidemic is its reproduction number, i.e., the average number of secondary cases caused by an infected individual. Again, since this metric is unrelated to temporal dynamics, we can derive it from S$\mathbb I$R model to simplify our calculation. According to the seminal work \cite{van2002reproduction}, the reproduction number is equivalent to the spectral radius of the next generation matrix $FV^{-1}$, where the $(m,n)$ element of matrix $F$ is the rate of new infections entering infected state $m$ caused by infected state $n$, and $(m,n)$ element of matrix $V$ is the rate at which infected state $m$ transfers to infected state $n$, assuming that the population remains near the disease-free equilibrium. In our edge-based S$\mathbb I$R model (\ref{theta2}) and (\ref{I2}), $\hat\xi_{ji}^{\mathbb I}$ can be treated as the infected state. Therefore, we have $M^2$ infected states in total. Recall that there should be $\frac{\hat\lambda_{ji}}{\hat\lambda_{ji}+\hat\gamma_{ji}}=T_{ji}$, where $T_{ji}$ is calculated from (\ref{Tij}). To simplify the derivations below, we further assume that $\hat\lambda_{ji}+\hat\gamma_{ji}=1$ (one can derive the same $R_0$ in (\ref{R0}) without this assumption, because $R_0$ is only related to $\frac{\hat\lambda_{ji}}{\hat\lambda_{ji}+\hat\gamma_{ji}}$). By differentiating (\ref{I2}) and taking (\ref{theta2}) into it, we have
\small
\begin{align}
&\dot{\hat\xi}_{ji}^{\mathbb I}(t)=- \hat\xi_{ji}^{\mathbb I}(t)+\notag\\
&\sum_{l=1}^M\sum_{\boldsymbol k} S_j(\boldsymbol k,0) \frac{k_i (k_l-\delta_{il}) p_j(\boldsymbol k)}{\overline k_{ji}} T_{lj} \hat\xi_{lj}^{\mathbb I}(t) \prod_{x\in[1,M]}\hat\theta^{k_x-\delta_{ix}-\delta_{xl}}_{xj}(t).\label{nonI3}
\end{align}
\normalsize

Then, to obtain the linearized subsystem for infected states about the disease-free equilibrium, we linearize (\ref{nonI3}) at the origin ($\hat\xi_{ji}^{\mathbb I}(t)=0$ and $\hat \theta_{ji}(t)=1$):
\begin{align}
&\dot{\hat\xi}_{ji}^{\mathbb I}(t)=- \hat\xi_{ji}^{\mathbb I}(t)+\sum_{l=1}^M\sum_{\boldsymbol k} S_j(\boldsymbol k,0) \frac{k_i (k_l-\delta_{il}) p_j(\boldsymbol k)}{\overline k_{ji}} T_{lj} \hat\xi_{lj}^{\mathbb I}(t).
\label{I3}
\end{align}

Then, we can construct $F$ as an $M^2\times M^2$ matrix, with $\big((i-1)M+j,(j-1)M+l\big)$ element equal to
\begin{gather}
\sum_{\boldsymbol k} S_j(\boldsymbol k,0)\frac{k_i (k_l-\delta_{il}) p_j(\boldsymbol k)}{\overline k_{ji}} T_{lj}, \quad \forall i,j,l\in [1,M],
\end{gather}
and construct $V$ as an $M^2\times M^2$ identity matrix. Reproduction number $R_0$ is therefore given by
\begin{gather}
R_0=\rho(FV^{-1})=\rho(F), \label{R0}
\end{gather}
where $\rho(\cdot)$ represents spectral radius. From Theorem 2 in \cite{van2002reproduction}, $R_0=1$ marks the epidemic threshold in the sense that the disease-free equilibrium is asymptotically stable if $R_0<1$, and is unstable if $R_0>1$. In particular, in the case of $S_j(\boldsymbol k,0)=1$ for all $j$ and $\boldsymbol k$, $R_0$ becomes the \textit{basic reproduction number}, i.e., the average number of secondary cases caused by an infected individual in a completely susceptible population. In this special case, the epidemic threshold $R_0=1$ is also in agreement with the threshold for multitype random network obtained by percolation theory\cite{allard2009heterogeneous,allard2015general}.
\section{Age-Stratified Vaccination\label{sect: vaccination}}
Under our proposed analytical framework, in this section, we present an age-stratified vaccination scheme and study its impact on epidemic outcomes. Considering a network with $N$ nodes of type $i$, we use the following function to characterize the immunization strategy for type-$i$ nodes\cite{shao2009structure,wang2014epidemic}:
\begin{gather}
\Phi_i(\tilde k_n)=\frac{\tilde k_n^{\alpha} } {\sum_{n=1}^{N}\tilde k_n^{\alpha}},-\infty<\alpha<+\infty, \label{immunization}
\end{gather}
where $\Phi_i(\tilde k_n)$ is the probability that a susceptible node $n$ of type $i$ with $\tilde k_n$-degree is chosen to be immunized, and $\alpha$ is an exponent quantifying the immunization preference towards nodes with high degree. We use the tilde operator on $k$ to represent the total degree of a node. A greater $\alpha$ indicates that nodes with higher degree (e.g., essential workers) are more likely to be immunized. In particular, $\alpha=\infty$ represents a node immunization process in an entire descending order, i.e., from the highest degree to the lowest degree. In contrast, when $\alpha=0$, we have $\Phi_i(\tilde k_n)=\frac{1}{N}$, implying a uniform immunization strategy for nodes of type $i$. Notice that since the full knowledge of a contact network is generally unavailable, immunizing nodes in a descending order is rather unrealistic. The value of $\alpha$ depends on the strategy and knowledge of a vaccine distributor.

Recall that $S_i(\boldsymbol k, 0)$ denotes the fraction of type-$i$ nodes with degree $\boldsymbol k$ that are initially susceptible. In our model, studying the impact of the immunization strategy in (\ref{immunization}) only requires solving new initial conditions $S_i^f(\boldsymbol k, 0)$, i.e., the fraction of type-$i$ nodes with degree $\boldsymbol k$ that are still susceptible when only $f$ fraction of type-$i$ nodes remain susceptible after implementing the immunization strategy, where $f = S_i(0) - v$, with $v$ being the fraction of type-$i$ nodes immunized by vaccination. Let $P_i(\tilde k)$ be the probability that a type-$i$ node is with degree $\tilde k$, $S^f_i(\tilde k, 0)$ and $A^f_i(\tilde k)$ be the fraction and the number of type-$i$ nodes with $\tilde k$ degree that are still susceptible when only $f$ fraction of type-$i$ nodes remain susceptible, respectively. Due to the fact that (\ref{immunization}) only depends on total node degree $\tilde k$, the subsequent development is only related to $\tilde k$ instead of the vector of node degree $\boldsymbol k$. According to the definitions, we have the relationship
\begin{gather}
P_i(\tilde k) S^f_i(\tilde k, 0) = \frac{A^f_i(\tilde k)}{N} \label{Pi}.
\end{gather}

After one susceptible node is immunized according to (\ref{immunization}), $A^f_i(\tilde k)$ becomes
\begin{gather}
A^{f-\frac{1}{N}}_i(\tilde k)=A^{f}_i(\tilde k)-\frac{P_i(\tilde k)S^f_i(\tilde k, 0)\tilde k^\alpha}{\overline {\tilde k^\alpha}(f)},
\label{Afi}
\end{gather}
where
\begin{gather}
\overline {\tilde k^\alpha}(f)\equiv\sum_{\tilde k} P_i(\tilde k)S^f_i(\tilde k, 0)\tilde k^\alpha.\label{kalpha}
\end{gather}

In the limit of $N\rightarrow\infty$, (\ref{Afi}) can be expressed as
\begin{gather}
\frac{d A^{f}_i(\tilde k)}{d f}=N \frac{P_i(\tilde k)S^f_i(\tilde k, 0)\tilde k^\alpha}{\overline {\tilde k^\alpha}(f)}.\label{deriv}
\end{gather}

Differentiating (\ref{Pi}) in terms of $f$ and plugging it into (\ref{deriv}), we obtain
\begin{gather}
\frac{d S^f_i(\tilde k, 0)}{d f}=\frac{S^f_i(\tilde k, 0)\tilde k^\alpha}{\overline {\tilde k^\alpha}(f)}. \label{deriv2}
\end{gather}

In the spirit of \cite{shao2009structure}, we define a new function $G_{\alpha}(x)=\sum_{\tilde k} P_i(\tilde k)S_i(\tilde k, 0)x^{\tilde k^\alpha}$ and introduce a new variable $t\equiv G^{-1}_{\alpha}(f)$ in order to solve (\ref{deriv2}), where $S_i(\tilde k, 0)$ in $G_{\alpha}(x)$ is the fraction of type-$i$ nodes with degree $\tilde k$ that are susceptible before implementing the immunization strategy. One can find that
\begin{gather}
S^f_i(\tilde k, 0)=t^{\tilde k^\alpha}S_i(\tilde k, 0), \label{removal1}
\end{gather}
exactly satisfies (\ref{deriv2}), which hence is the solution to (\ref{deriv2}). Equivalently, considering the vector of node degree $\tilde{\boldsymbol k}$ with $\tilde k$ as the total degree, we have
\begin{gather}
S^f_i(\tilde{\boldsymbol k}, 0)=t^{\tilde k^\alpha}S_i(\tilde{\boldsymbol k}, 0), \label{removal2}
\end{gather}
By simply replacing $S_i(\tilde{\boldsymbol k}, 0)$ with the new initial conditions $S^f_i(\tilde{\boldsymbol k}, 0)$, we can obtain the various epidemic outcomes in the preceding section by taking the age-stratified immunization into account.

\section{Simulations\label{sect: simulation}}
\subsection{Parameter settings\label{subsect: settings}}
We now study the impact of control policies, particularly age-specific vaccination strategies, for the COVID-19 pandemic. We use the estimated social contact data by age groups for the United States to conduct our simulations, where $C=[c_{i j}]$ represents the contact matrix by age, with $c_{i j}$ being the average number of contacts that a node of type $i$ has with nodes of type $j$\cite{prem2017projecting}. Notice, however, that the conclusions drawn from our simulations are also generalizable to many other countries, because age-stratified contact exhibits similar patterns in most countries\cite{mossong2008social,prem2017projecting}. The population is partitioned into $M=6$ age groups, i.e., populations of $[0, 4]$, $[5, 19]$, $[20, 39]$, $[40, 49]$, $[50, 59]$ and $60+$ years old.

Following \cite{davies2020age}, we assume that the susceptibility to infection for adults over 20 years is identical, and the susceptibility to infection for individuals under 20 years old is half of that for adults over 20 years old. Specifically, we set transmission rate $\lambda^I_{ij}= \lambda$ for all $i\in [1,6]$ and $j\in[3,6]$, and $\lambda^I_{ij}= \frac{1}{2}\lambda$ for all $i\in [1,6]$ and $j\in[1,2]$. We set $\lambda^A_{ij}= \frac{2}{3}\lambda^I_{ij}$ to account for the fact that asymptomatic people are less infectious than symptomatic ones\footnote{It is commonly recognized that symptomatic patients are more infectious than asymptomatic ones because cough and sneeze could help spread the virus.}. The course of an epidemic is primarily governed by the basic reproduction number. Being consistent with \cite{britton2020mathematical}, we assume that the basic reproduction number is $R_0=2.5$, and derive the transmission rate $\lambda$ from (\ref{R0}) accordingly. Given that young people develop milder symptoms or no symptoms more frequently than the elderly, the symptomatic probabilities are set to $20\%$, $20\%$, $30\%$, $40\%$, $50\%$, and $60\%$, and the probabilities of needing to be hospitalized for symptomatic cases are set to $0.10\%$, $0.23\%$, $2.19\%$, $4.90\%$, $10.20\%$, and $20.82\%$ from young age groups to old age groups\cite{ferguson2020report}. Furthermore, the infection fatality ratios are set to $0.003\%$, $0.01\%$, $0.06\%$, $0.16\%$, $0.60\%$, and $3.64\%$ from the young to the elderly, respectively\cite{ferguson2020report}. We set the average time from the exposure to the onset of being infected (i.e., A or I states) to 5 days, the average infection period to $7$ days if not admitted to hospital, and the average time stay in hospitals to $10$ days\cite{wang2020clinical}. To compare the effectiveness of vaccination prioritization strategies, we consider a completely susceptible population before vaccination. Later, this assumption is relaxed in Fig. \ref{fig: MortalityImm} to demonstrate the consistency of our conclusion by considering a population with a high level of natural immunity.

\begin{figure}[t]
\centering
\includegraphics[width=2.5in]{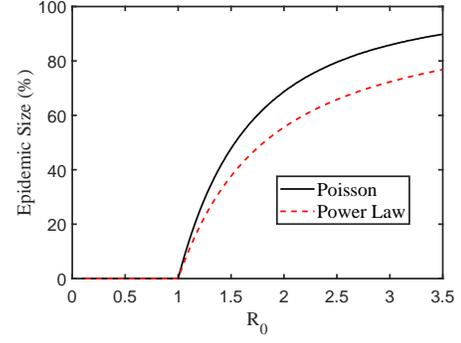}
\caption{Epidemic size versus reproduction number $R_0$.}
\label{fig: powerlaw}
\end{figure}

\subsubsection{Modeling of Universal Masking}
At the early stage of vaccine distribution campaign, masking and/or social distancing measures are still needed. Thus, assessing the effectiveness of an vaccination prioritization strategy requires the considerations of ongoing NPI policies. We denote the population contact matrix as $C=C^{h}+C^{w}+C^{s}+C^{o}$, where $C^{h}$, $C^{w}$, $C^{s}$ and $C^{o}$ are the age-stratified contact matrices for home, workplace, school, and other locations, respectively\cite{prem2017projecting}. For instance, the $(i,j)$-th element in $C^{h}$, denoted by $c^h_{i j}$, is the average degree from a node of type $i$ to nodes of type $j$ at home. By considering the NPIs, we modify the population contact matrix $C$ as follows.
\begin{equation}
C=C^{h}+g\big(C^{w}+C^{s}+C^{o}\big),\label{matrix}
\end{equation}
where $g\in [0,1]$ is the scaling factor accounting for the change in transmission rate per contact due to the presence of NPIs. We assume that widespread face masking in public (i.e., workplace, school, and other locations) are recommended. Since reducing the transmission rate for an edge (contact) by $1-g$ is equivalent to removing this edge with probability $1-g$ for the spread of epidemic\cite{meyers2005network}, we directly scale the contact matrices $C^{w}$, $C^{s}$, and $C^{o}$ to reflect the reduction in transmission rates in these places. The value of $g$ can be estimated from mask coverage (the fraction of population wearing masks) and mask efficacy (the fraction of effective transmissions blocked by masking)\cite{eikenberry2020mask}. In what follows, we assume $g=0.3$ for illustrative purpose. By this scaling, reproduction number $R_0$ is pushed from $2.5$ to $1.16$. We will compare different vaccination prioritization strategies under no-masking scenario with $g=1$ and masking scenario with $g=0.3$. We remark that weaker mask use in conjunction with other NPIs in public places (i.e., social distancing measures) may have a similar effect on $g$. This is because social distancing also reduces the transmission opportunity between two individuals, which has no difference with masking in terms of mathematical modeling.

To demonstrate that our parameter $g=0.3$ is realistic, we refer the readers to the reference \cite{eikenberry2020mask}. According to Ref. \cite{eikenberry2020mask}, when the product of mask coverage and mask efficacy is $0.6$, e.g., mask coverage is $0.75$ and mask efficacy is $0.8$, the relative transmission rate of COVID-19 reduces to $0.3$ compared with the no-masking case. In the real world, the efficacy of surgical masks is estimated to be about $0.8$\cite{eikenberry2020mask}. Therefore, three quarters of population wearing surgical masks in the public places (i.e., workplace, school, and other locations) may lead to $g=0.3$.

\subsubsection{Impact of Population Structure Heterogeneity}
Even with the same $R_0$ and contact matrix $C$ (containing average contact number $c_{ij}$ across age groups), the epidemic may still spread differently in networks because of the assumption on degree distributions. Fig. \ref{fig: powerlaw} sheds light on the effects of structural heterogeneity on epidemic outcome. Based on the same contact matrix, we examine two types of degree distributions: Poisson distribution and power law distribution with the law's exponent equal to $2.5$\cite{meyers2005network}. From the estimates in \cite{prem2017projecting}, the contact numbers $c_{ij}$ are assumed to follow Poisson distributions. Compared with Poisson distributions, power law distributions have a quite ``heterogeneous'' structure: it contains not only many nodes with few contacts, but also a handful of ``superspreaders'' with very high degree. As illustrated in Fig. \ref{fig: powerlaw}, power law network shrinks the epidemic size compared with Poisson network, which clearly reveals that the same $R_0$ and matrix of average contact number are still not enough to accurately forecast epidemic dynamics. In fact, the assumption that contact networks follow Poisson distributions is rather ideal, as it fails to capture the superspreader events that may greatly drive the transmissions of COVID-19\cite{tkachenko2020persistent}. An estimate of degree distributions of real-world contact networks is still needed in the future research to improve the precision of projected epidemic results. However, noting that our mission in this section is to seek the effective vaccination prioritization policies rather than providing the exact or even best estimate of epidemic dynamics, we assume that the contact network follows Poisson distributions without loss of generality.

\subsection{Effectiveness of Vaccination Prioritization Strategies}
\begin{figure}
\centering
\begin{subfigure}[b]{0.49\linewidth}
\centering
\includegraphics[width=1.9in]{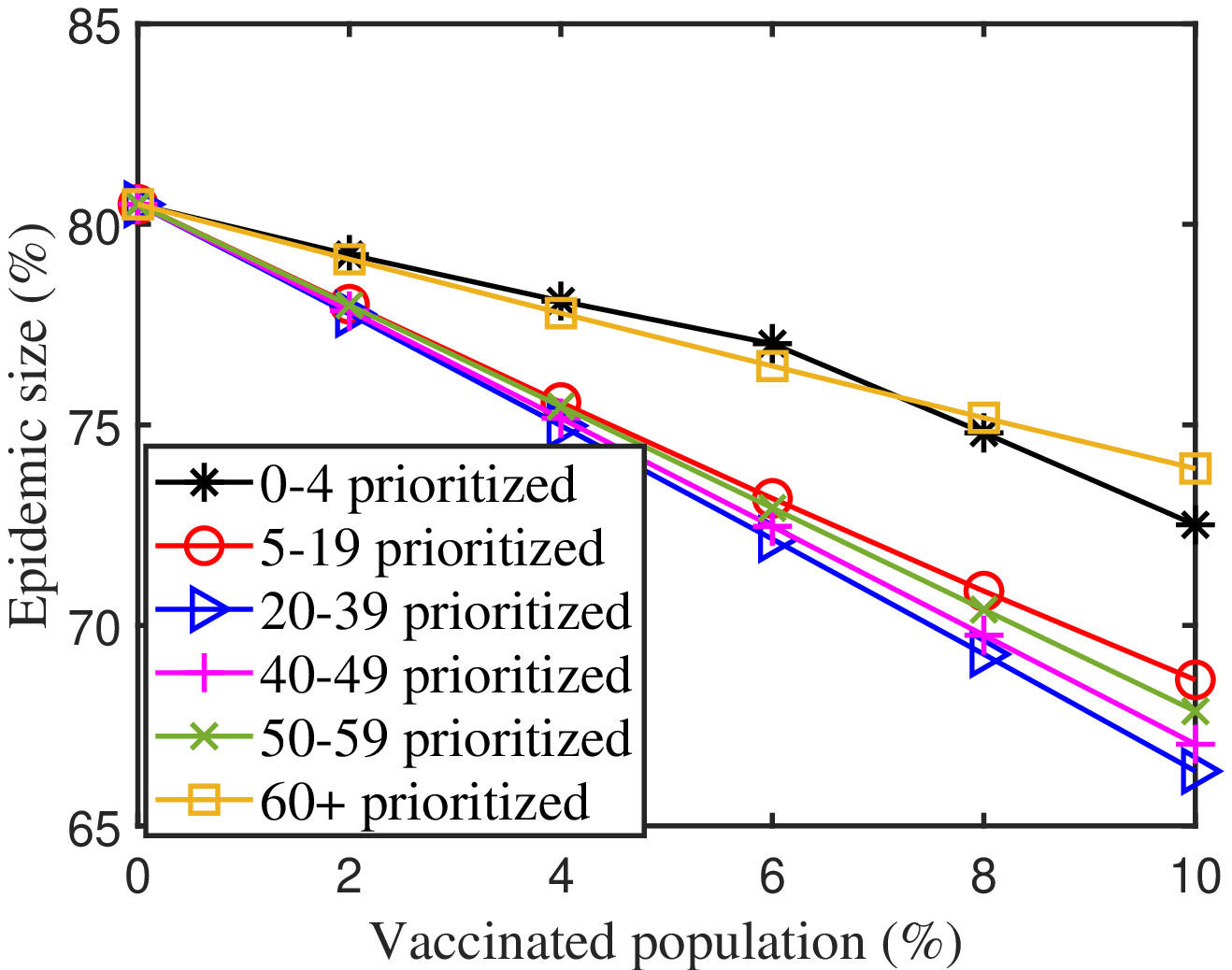}
\caption{No-masking scenario ($g=1$).}
\label{fig: InfectedPre0}
\end{subfigure}
\hfill
\begin{subfigure}[b]{0.49\linewidth}
\centering
\includegraphics[width=1.9in]{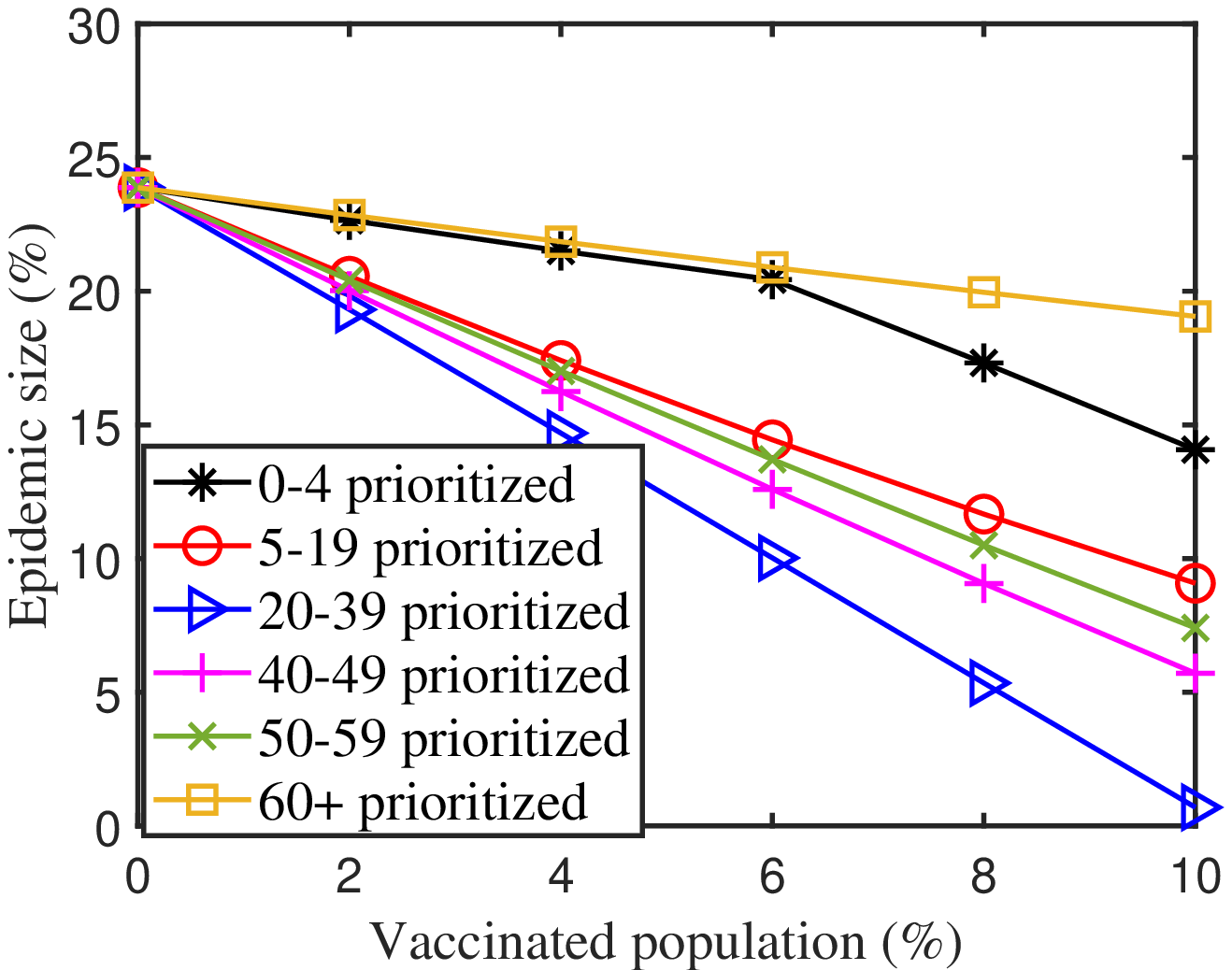}
\caption{Masking scenario ($g=0.3$).}
\label{fig: InfectedPost0}
\end{subfigure}
\caption{Epidemic size versus different immunization prioritization strategies.}
\label{fig: Infected}
\end{figure}

\begin{figure}
\begin{subfigure}[b]{0.49\linewidth}
\centering
\includegraphics[width=1.9in]{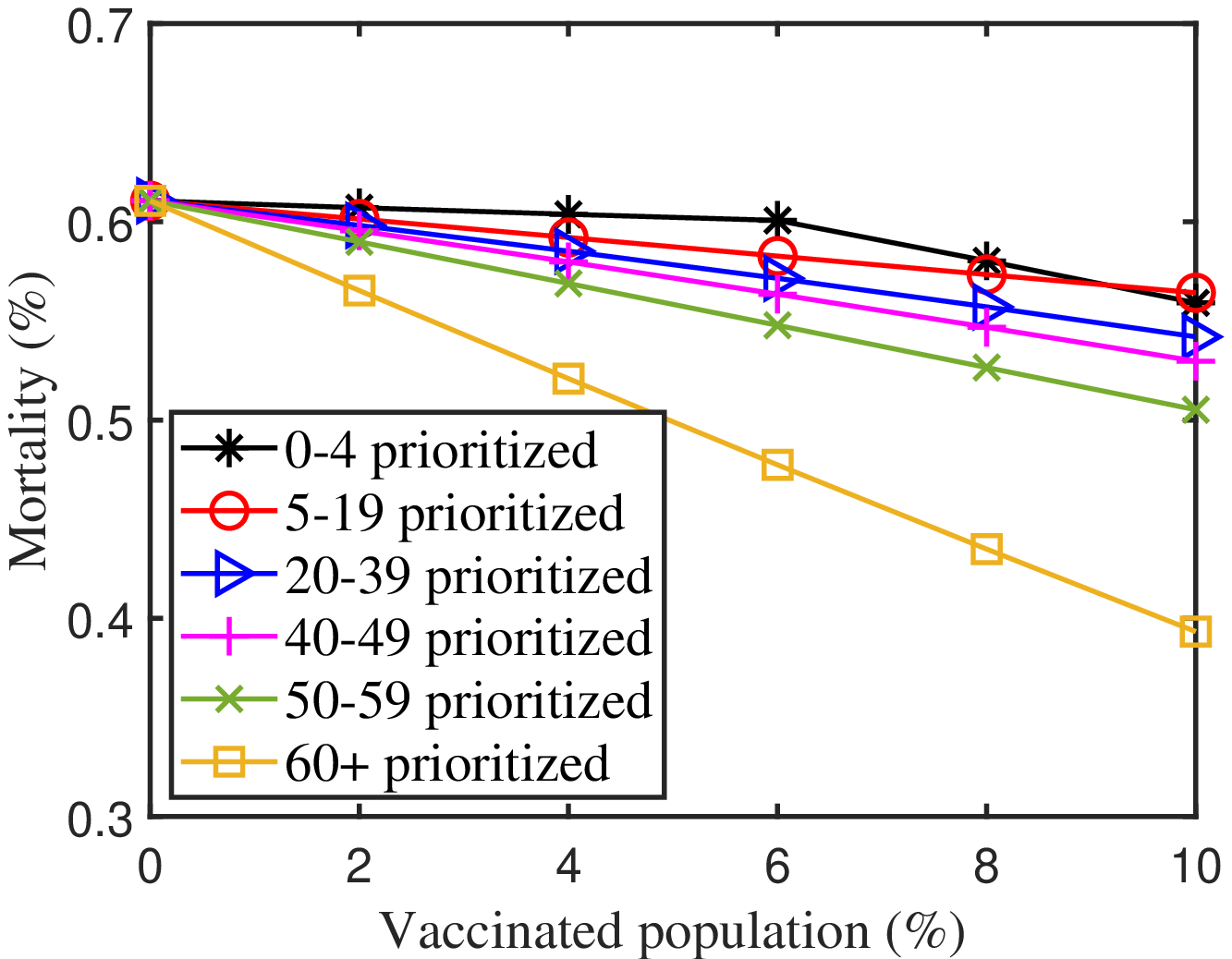}
\caption{No-masking scenario ($g=1$).}
\label{fig: MortalityPre0}
\end{subfigure}
\hfill
\begin{subfigure}[b]{0.49\linewidth}
\centering
\includegraphics[width=1.9in]{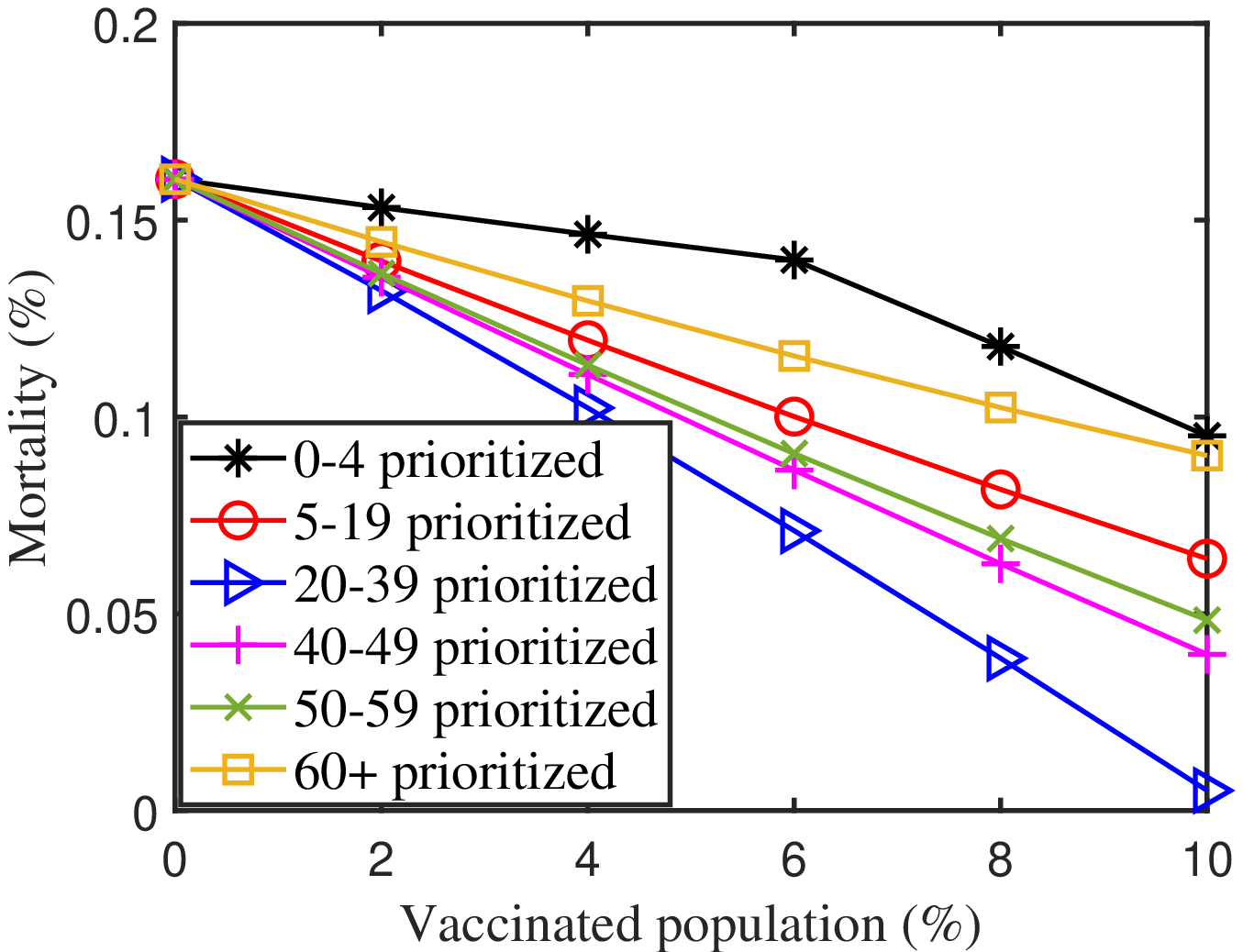}
\caption{Masking scenario ($g=0.3$).}
\label{fig: MortalityPost0}
\end{subfigure}
\caption{Mortality versus different immunization prioritization strategies.}
\label{fig: Mortality}
\end{figure}

We intend to compare the effectiveness of age-specfic vaccination prioritization strategies under two scenarios: no-masking scenario and masking scenario. Unless specified otherwise, we consider uniform vaccination within the same age group by setting $\alpha=0$ in (\ref{removal2}). Although we conduct simulations by assuming the basic reproduction number $R_0=2.5$, our conclusions below also hold for other reasonable estimated $R_0$ values for COVID-19, such as from $2$ to $3.5$. We assume that the vaccine efficacy for individuals below $65$ years old is $95.6\%$, and for individuals over $65$ years old is $86.4\%$ according to the Moderna's clinical trial data\cite{Moderna}. In fact, our findings below also hold for Pfizer's vaccine efficacy, which is about $95\%$ for all age groups\cite{polack2020safety}. Since the elderly aged 60+ has the lowest average contact number but the highest mortality and illness severity, while the adults aged $20-39$ have a high average contact rate and low mortality and illness severity, we call the adults aged $20-39$ as the \textit{high-transmission} group, and the elderly aged 60+ as the \textit{high-risk} group. We remark that, even though the children aged 5-19 have the highest average contact number among the population\cite{prem2017projecting}, they are assumed to be less susceptible to the infection as mentioned before, thus contributing less to the COVID-19 transmissions than the adults aged $20-39$.

In Fig. \ref{fig: Infected}, we compare the epidemic size under different vaccination prioritization strategies by varying the fraction of the whole population being vaccinated.
In the figure, we use ``20-39 prioritized'' (or other age ranges) to represent that the vaccine doses are all given to the population aged 20-39. In particular, since the children aged $0-4$ only constitute about $6\%$ of the whole population, the remaining vaccines, if all the children aged $0-4$ get vaccinated (i.e., in the case where $8\%$ or $10\%$ population is vaccinated in the figure), are uniformly allocated to other age groups. As shown in the figure, prioritization of the adults aged 20-39 is most effective in blocking the transmissions and reducing the infections under both no-masking and masking scenarios. However, we should notice the difference: the reduction in epidemic size achieved by prioritization of the high-transmission group under masking scenario is much more significant than that of no-masking scenario. As illustrated in Fig. \ref{fig: powerlaw}, when around $R_0=2.5$, i.e., under no-masking scenario, epidemic size decreases slowly with $R_0$. As a result, no matter which vaccination strategy is implemented, it will not affect the epidemic size much, as observed from Fig. \ref{fig: Infected}(a). In contrast, under masking scenario with $R_0=1.16$, epidemic size shrinks fast as $R_0$ reduces. For this reason, prioritizing the high-transmission group reduces the epidemic size significantly as shown in Fig. \ref{fig: Infected}(b).

In Fig. \ref{fig: Mortality}, we investigate which age-specific vaccination prioritization strategy reduces the mortality (the death toll over the whole population) most. This metric is calculated from the epidemic size and the age-dependent mortality ratios. As shown in Fig. \ref{fig: Mortality}(a), prioritizing the elderly achieves the lowest mortality under no-masking scenario. This is due to two facts: 1) no matter which prioritization strategy is implemented, limited vaccine doses will not decrease the epidemic size much when $R_0$ is great. 2) The elderly people have remarkably higher mortality risk than the remaining population. Consequently, protecting the elderly directly is a wise method in such a case. Nonetheless, under masking scenario, inoculating the adults aged $20-39$ becomes the most effective strategy to reduce the mortality as illustrated in Fig. \ref{fig: Mortality}(b). This is because prioritization of high-transmission groups substantially blocks COVID-19 transmissions in small $R_0$ region as aforementioned, thereby in turn protecting the elderly even though they have not been vaccinated. There exist some related simulation studies in the literature. It is shown in \cite{jentsch2020prioritising} that vaccinating high-transmission group is the best strategy to minimize the death from COVID-19 when the NPIs (for Ontario) is combined with high level of natural immunity. Our results further illustrate that relatively strong NPI strategies, even without natural immunity, could lead to the same conclusion. In \cite{matrajt2020vaccine}, the researchers show that high-transmission group should be prioritized to minimize death only when the vaccination covers a large proportion of the population (e.g., over $40\%$ coverage when the vaccine efficacy is $100\%$). This finding may be due to that they have not combined vaccination with NPIs. As a consequence of this difference, our results instead indicate that the high-transmission group should be prioritized under the presence of relatively strong NPIs, even if the vaccine coverage is very limited, say, $2\%$, as shown in Fig. \ref{fig: Mortality}(b).

\begin{figure}
\begin{subfigure}[b]{0.49\linewidth}
\centering
\includegraphics[width=1.9in]{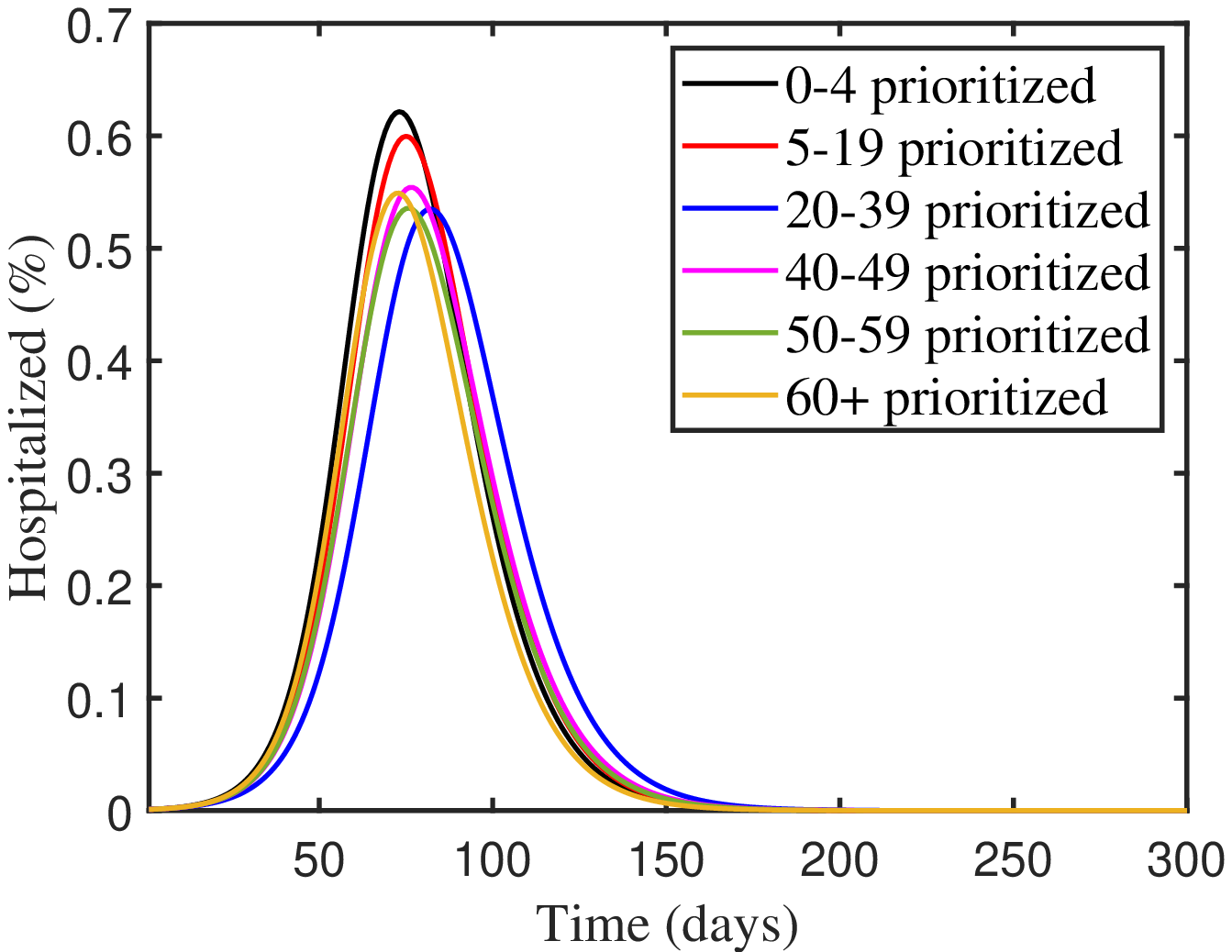}
\caption{5\% population vaccinated under no-masking scenario ($g=1$).}
\label{fig: HospPre0}
\end{subfigure}
\hfill
\begin{subfigure}[b]{0.49\linewidth}
\centering
\includegraphics[width=1.9in]{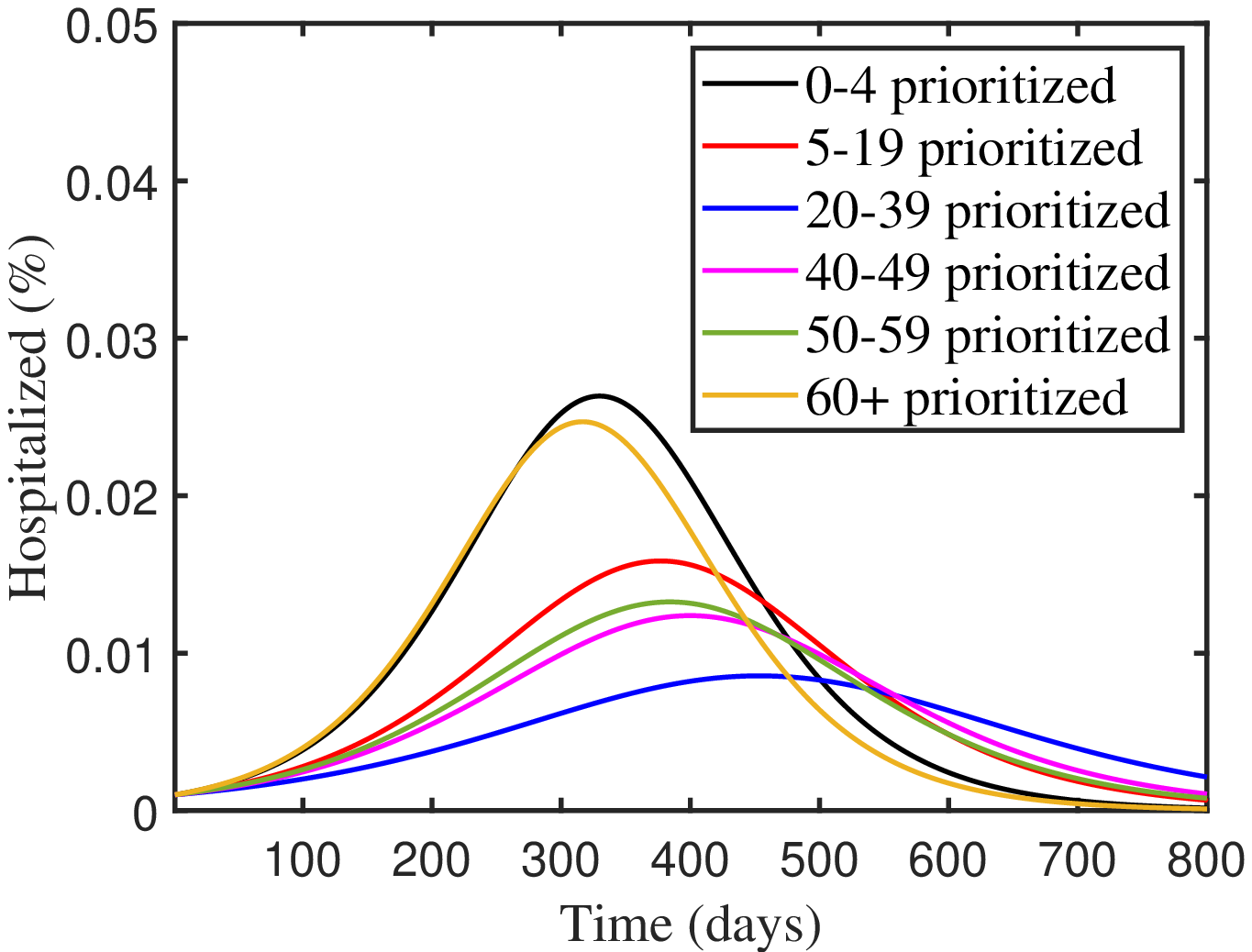}
\caption{5\% population vaccinated under masking scenario ($g=0.3$).}
\label{fig: HospPost0}
\end{subfigure}
\caption{Hospitalizations versus different immunization prioritization strategies.}
\label{fig: Hosp}
\end{figure}

In Fig. \ref{fig: Hosp}, we evaluate how different vaccination prioritization strategies affect hospitalizations. Similar to the results for mortality, vaccinating high-transmission group first is still more effective under masking scenario, as shown in Fig. \ref{fig: Hosp}(b). On the other hand, while both severity and mortality risks increase with age, the disparity in severity ratio between the elderly and younger groups is not as significant as the disparity in the mortality ratio. As a result, under no-masking scenario,  vaccinating the adults aged $20-39$ or $50-59$ first even slightly outperforms vaccinating the elderly first in terms of reducing hospitalizations from COVID-19 as depicted in Fig. \ref{fig: Hosp}(a), because the former age groups have much higher average contact rates than the elderly people.

\begin{figure}
\centering
\begin{subfigure}[b]{0.49\linewidth}
\centering
\includegraphics[width=1.9in]{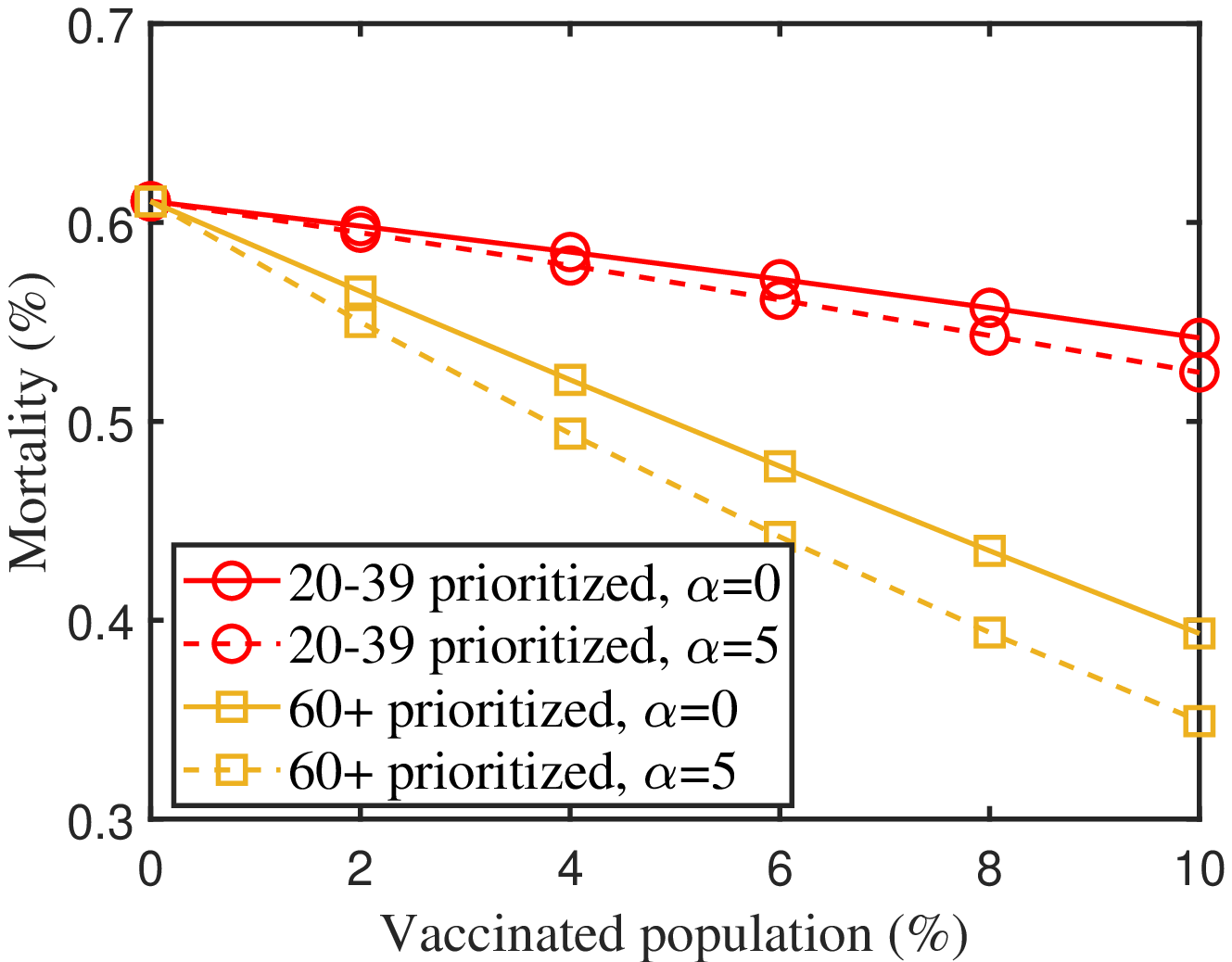}
\caption{No-masking scenario ($g=1$).}
\label{fig: MortalityAlpha0}
\end{subfigure}
\hfill
\begin{subfigure}[b]{0.49\linewidth}
\centering
\includegraphics[width=1.9in]{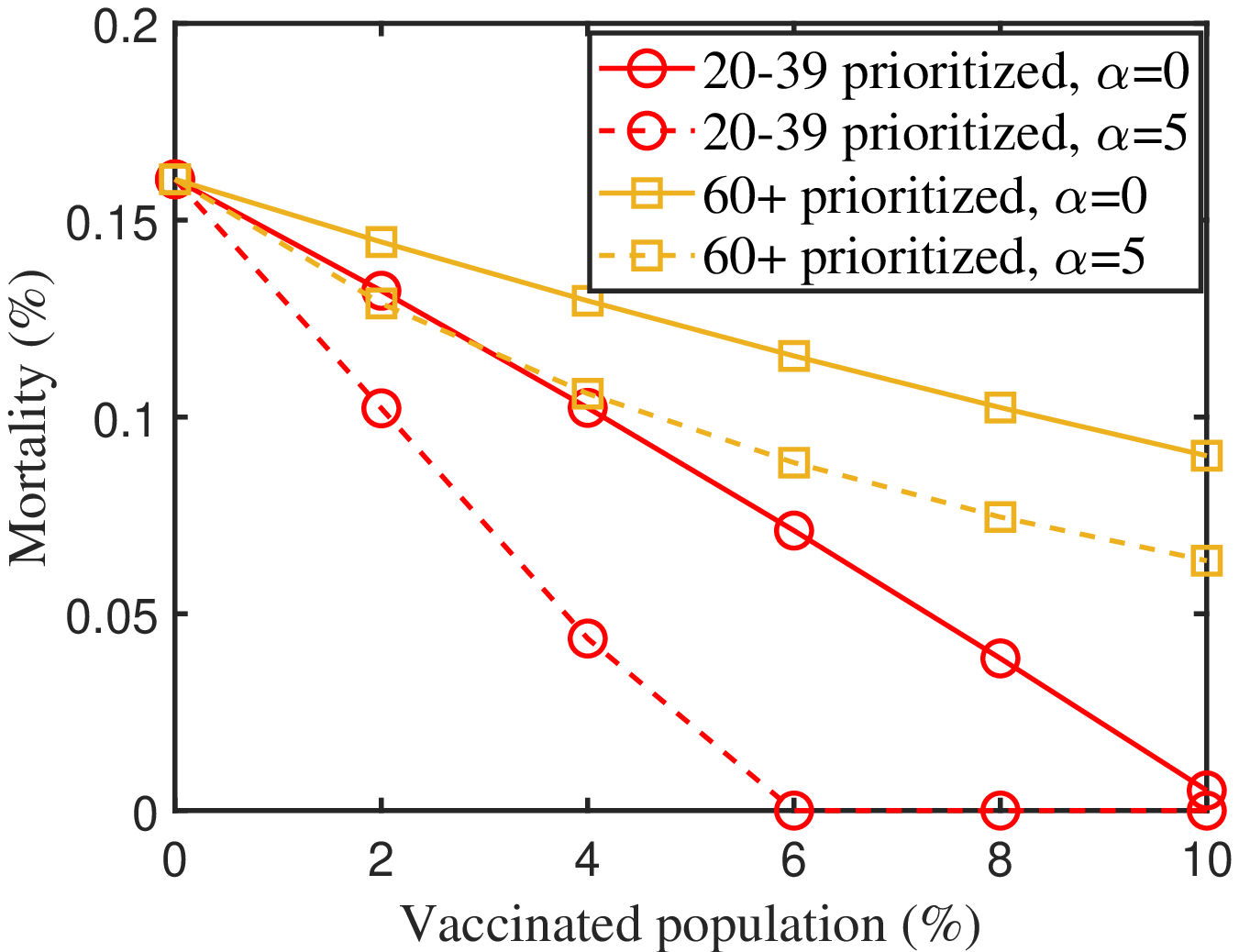}
\caption{Masking scenario ($g=0.3$).}
\label{fig: MortalityAlpha5}
\end{subfigure}
\caption{Mortality versus different immunization prioritization strategies with different $\alpha$.}
\label{fig: Alpha}
\end{figure}

Fig. \ref{fig: Alpha} evaluates the performance of different age-specific vaccination strategies versus $\alpha$. (\ref{removal2}) with $\alpha>0$ corresponds to a vaccine plan targeting at people with high activity level (e.g., essential workers) in that age group. Thus, it is not surprising to see that the vaccination strategies with $\alpha=5$ outperform the corresponding vaccination strategies with $\alpha=0$ in both no-masking and masking scenarios. The effect of changing $\alpha$ is more significant in Fig. \ref{fig: Alpha}(b) than Fig. \ref{fig: Alpha}(a), as the epidemic results in small $R_0$ region are more sensitive to the limited vaccination. In real world, $\alpha$ reflects the vaccine allocation plan and the knowledge of the vaccine distributor towards the population structure, which must be taken into account to forecast the effectiveness of certain immunization strategies. This preferential immunization for human networks with general degree distributions, however, cannot be evaluated based on traditional age-stratified homogeneous-mixing models.

\begin{figure}
\begin{subfigure}[b]{0.49\linewidth}
\centering
\includegraphics[width=1.9in]{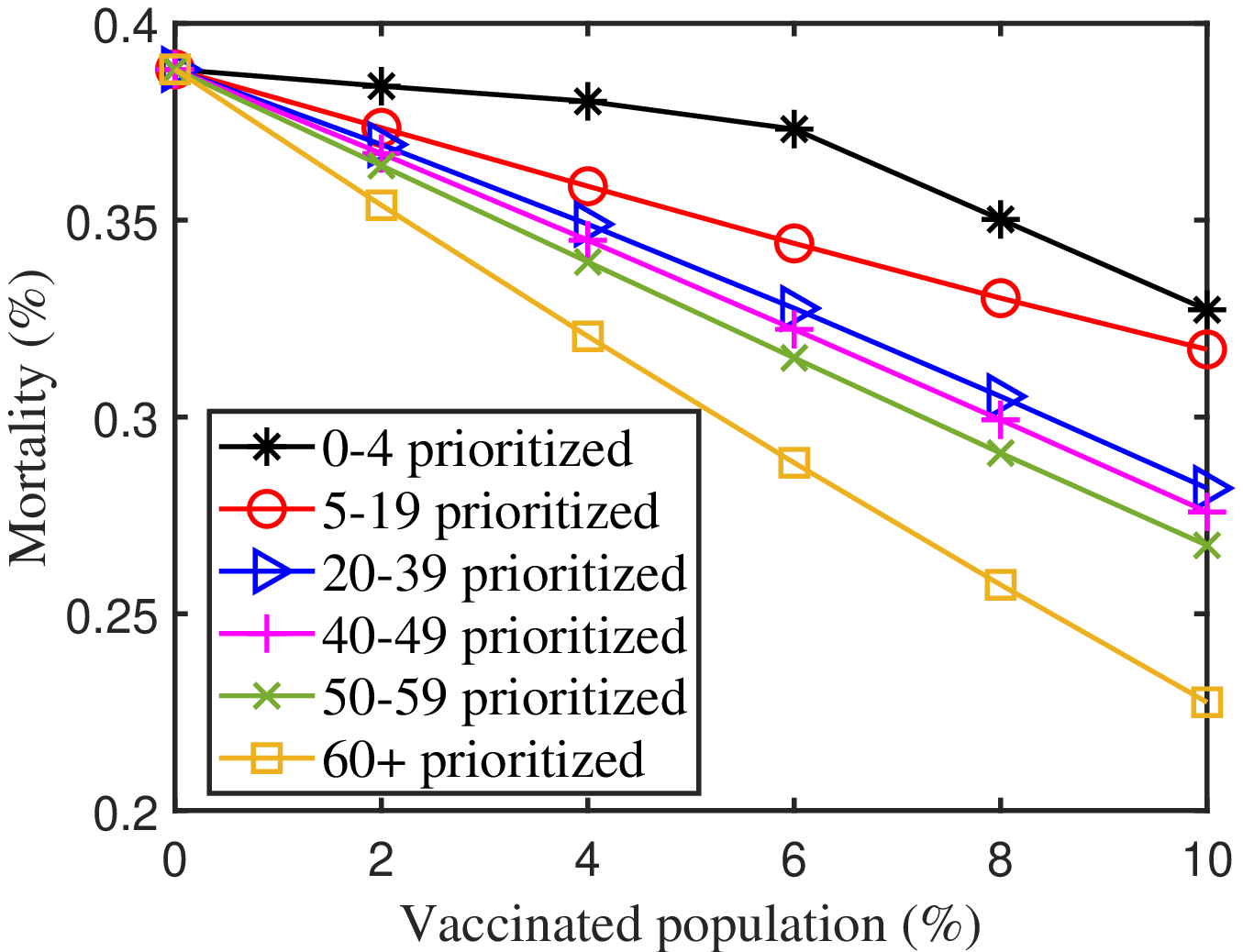}
\caption{No-masking scenario ($g=1$)}
\label{fig: MortalityPre5Imm}
\end{subfigure}
\hfill
\begin{subfigure}[b]{0.49\linewidth}
\includegraphics[width=1.9in]{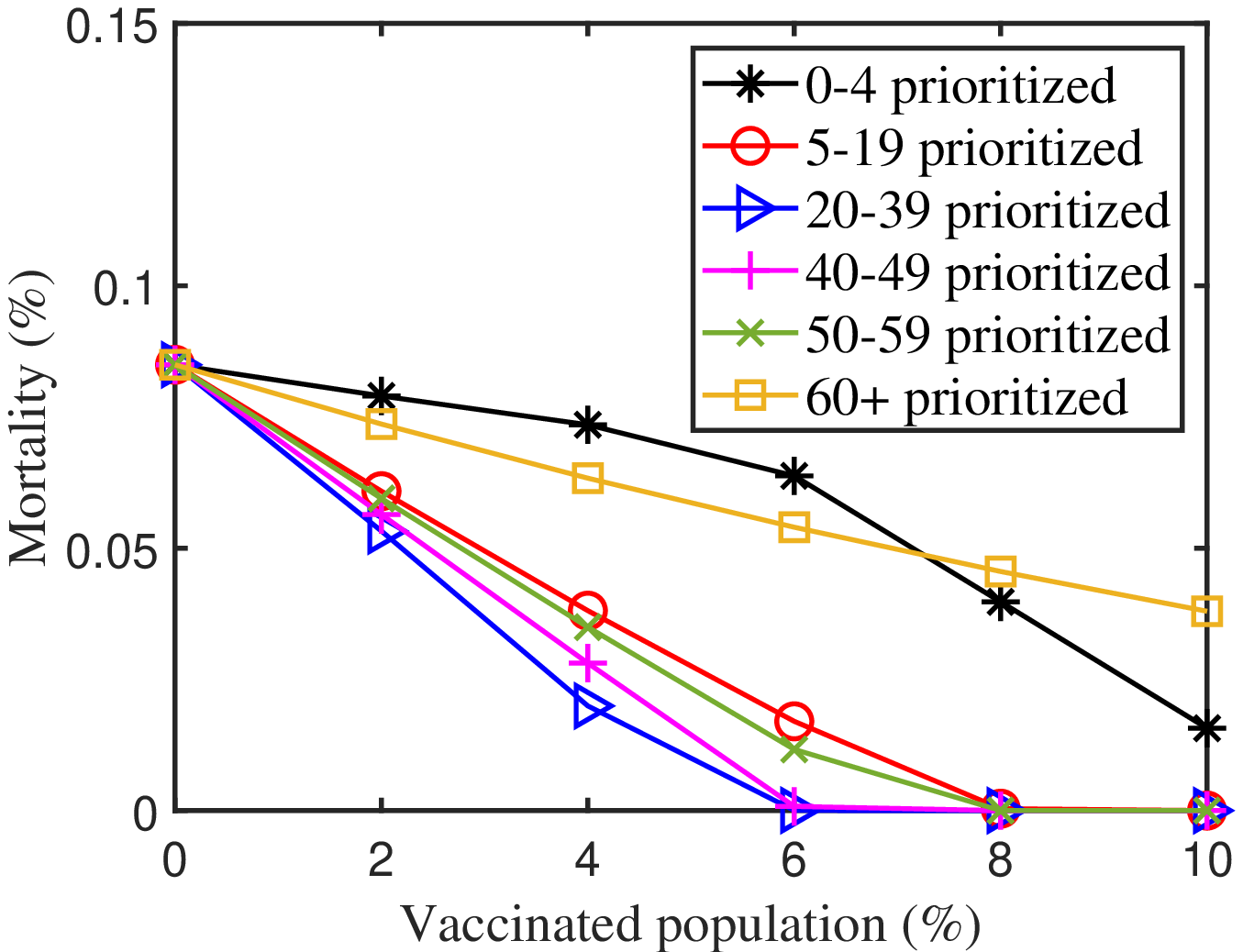}
\caption{Masking scenario ($g=0.5$).}
\label{fig: MortalityPost5Imm}
\end{subfigure}
\caption{Mortality versus different immunization prioritization strategies with $20\%$ population naturally immune before vaccination.}
\label{fig: MortalityImm}
\end{figure}

Fig. \ref{fig: MortalityImm} examines the consistency of our conclusions under a hard-hitting scenario with $20\%$ population naturally immune before vaccination. To obtain the initial conditions $S_i(\boldsymbol k,0)$ for this case, we simulate the disease spread according to the time-dependent dynamics in (\ref{d1})-(\ref{d6}) until when around $20\%$ people get infected. Due to the natural immunity, $R_0$ reduces from $2.5$ to $1.78$ under the no-masking scenario. To sustain $R_0$ above $1$, we set $g=0.5$, which corresponds to a looser masking measure, resulting in $R_0=1.1$ under the masking scenario. As can be observed from Fig. \ref{fig: MortalityImm}, vaccinating the elderly still reduces the mortality from COVID-19 most in the no-masking case, and vaccinating adults aged $20-39$ still decreases the mortality most in the masking scenario. This phenomenon demonstrates that our conclusion still holds when a high level of natural immunity has already been achieved.

\subsection{Epidemic Probability}
\begin{figure}[t]
\centering
\includegraphics[width=2.5in]{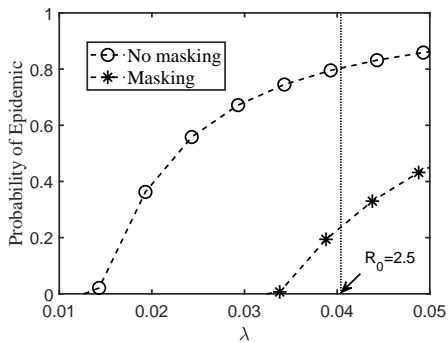}
\caption{Epidemic probability versus transmission rate $\lambda$ under either no-masking or masking case. The vertical line marks the value of $\lambda$ corresponding to $R_0=2.5$ under no-masking scenario.}
\label{fig: Prob}
\end{figure}

The proposed epidemic model is capable of capturing the stochastic property of epidemics. For countries or areas that have no active cases inside, it is useful to estimate the probability that a new import case, if not quarantined properly, sparks an epidemic. In Fig. \ref{fig: Prob}, we study how universal masking could suppress the epidemic probability under both no-masking and masking scenarios. Recall that the epidemic probability is the likelihood that a zero patient randomly chosen from the population starts an epidemic. Under the no-masking scenario, it is shown that the epidemic probability is about $0.8$ when $R_0= 2.5$, implying that most new cases, if not isolated, will give rise to an epidemic. Conversely, face mask wearing effectively suppresses the epidemic probability to about $0.25$. As reported in China, import cold-chain food contamination is even a source for COVID-19 resurgence, and therefore it is nearly impossible to isolate all new (or import) case properly. Consequently, taking some low-cost control policies, such as universal masking, is still important for disease-free areas to reduce or eliminate the risk of COVID-19 outbreak or resurgence.

\section{Conclusion\label{sect: conclusion}}
To combat the COVID-19 pandemic, one of the most important research tasks is to find out how to effectively decrease mortality and severe illness from COVID-19. To achieve this goal, we present a unified analytical framework for COVID-19 by considering both age-dependent risks and heterogeneity in contact networks within and across age groups. Under this framework, we use a novel age-stratified SEAHIR compartmental model to account for the distinct dynamics in a micro-state level, and employ the multitype random network approach to characterize the spread of epidemics. Several critical epidemiological metrics, including time-dependent dynamics, epidemic size, epidemic probability, and reproduction number are rigorously derived to capture essential features to be used to manage the pandemic.

Based on our proposed epidemic model, we have also studied the vaccination problem. It turns out that what is the best vaccination prioritization strategy to decrease mortality and hospitalizations depends on the reproduction number $R_0$. In other words, the effective strategies may vary across different areas, and heavily depends on the level of local NPI policies, such as masking, that suppress COVID-19 transmissions. We conclude that vaccinating the high-risk group is only effective in reducing mortality when $R_0$ is relatively high, e.g., under the no-masking scenario, whereas vaccinating the high-transmission group turns out to be the wise strategy if intervention policies have already suppressed $R_0$ at a low level. Although there are many social and ethical considerations in vaccination allocation, our results provide the rationale for vaccination prioritization at early stage of vaccination campaign.

There are several promising directions for future research. First, the COVID-19 reinfection can be incorporated into the epidemic model. In this paper, we assume that once a person becomes immune (either via getting infected or vaccinated), the person will never contract the disease. In a relatively short term (e.g., several months), this assumption may be reasonable given the rare reports of reinfection and the current understanding of the vaccination. However, how long the protective antibodies last remains an open problem. To take the possible reinfection into account, we need to break the state $R$ into recovery, vaccinated and death states, and consider the transition from the recovery and vaccinated states to state $S$. Under this case, we can still obtain the time-dependent epidemic dynamics by using the proposed approach in Section \ref{sect: dynamics}. Nevertheless, since the final state of the network may not be disease-free due to the existence of reinfection, other fundamental epidemic metrics, such as epidemic size and reproduction number, cannot be calculated via the proposed approach, which is worth studying in the future. Second, it is useful to evaluate many other NPIs based on our proposed model. For instance, what is the impact of limiting some gatherings or events, such as mass gatherings in bars, gyms, and churches, on the epidemic spread? To answer these kinds of questions, one can simulate a realistic contact network (e.g., with households, schools, bars, gyms, and churches), as in \cite{meyers2005network}, to discover the impact of different gatherings or events on the network structure, and then perform the epidemic analysis using our mathematical framework by removing the edges contributed by these gatherings.

\bibliographystyle{IEEEtran}
\bibliography{mybibtex}

\begin{thebibliography}{10}
\providecommand{\url}[1]{#1}
\csname url@samestyle\endcsname
\providecommand{\newblock}{\relax}
\providecommand{\bibinfo}[2]{#2}
\providecommand{\BIBentrySTDinterwordspacing}{\spaceskip=0pt\relax}
\providecommand{\BIBentryALTinterwordstretchfactor}{4}
\providecommand{\BIBentryALTinterwordspacing}{\spaceskip=\fontdimen2\font plus
\BIBentryALTinterwordstretchfactor\fontdimen3\font minus
  \fontdimen4\font\relax}
\providecommand{\BIBforeignlanguage}[2]{{%
\expandafter\ifx\csname l@#1\endcsname\relax
\typeout{** WARNING: IEEEtran.bst: No hyphenation pattern has been}%
\typeout{** loaded for the language `#1'. Using the pattern for}%
\typeout{** the default language instead.}%
\else
\language=\csname l@#1\endcsname
\fi
#2}}
\providecommand{\BIBdecl}{\relax}
\BIBdecl

\bibitem{WHO}
{World Health Organization (WHO)}, ``{WHO Coronavirus Disease (COVID-19)
  Dashboard},'' https://covid19.who.int/, 2020.

\bibitem{CDC}
{Centers for Disease Control And Prevention (CDC)}, ``Similarities and
  differences between flu and {COVID-19},''
  https://www.cdc.gov/flu/symptoms/flu-vs-covid19.htm, 2020.

\bibitem{ferguson2020report}
N.~Ferguson, D.~Laydon, G.~Nedjati~Gilani, N.~Imai, K.~Ainslie, M.~Baguelin,
  S.~Bhatia, A.~Boonyasiri, Z.~Cucunuba~Perez, G.~Cuomo-Dannenburg
  \emph{et~al.}, ``{Report 9: Impact of non-pharmaceutical interventions (NPIs)
  to reduce COVID19 mortality and healthcare demand},''
  https://dsprdpub.cc.ic.ac.uk:8443/handle/10044/1/77482, 2020.

\bibitem{verity2020estimates}
R.~Verity, L.~C. Okell, I.~Dorigatti, P.~Winskill, C.~Whittaker, N.~Imai,
  G.~Cuomo-Dannenburg, H.~Thompson, P.~G. Walker, H.~Fu \emph{et~al.},
  ``{Estimates of the severity of coronavirus disease 2019: a model-based
  analysis},'' \emph{The Lancet infectious diseases}, June 2020.

\bibitem{singh2020age}
R.~Singh and R.~Adhikari, ``{Age-structured impact of social distancing on the
  COVID-19 epidemic in India},'' \emph{arXiv preprint arXiv:2003.12055}, 2020.

\bibitem{balabdaoui2020age}
F.~Balabdaoui and D.~Mohr, ``{Age-stratified model of the COVID-19 epidemic to
  analyze the impact of relaxing lockdown measures: nowcasting and forecasting
  for Switzerland},'' \emph{medRxiv}, 2020.

\bibitem{tuite2020mathematical}
A.~R. Tuite, D.~N. Fisman, and A.~L. Greer, ``{Mathematical modelling of
  COVID-19 transmission and mitigation strategies in the population of Ontario,
  Canada},'' \emph{CMAJ}, vol. 192, no.~19, pp. 497--505, May 2020.

\bibitem{jentsch2020prioritising}
P.~Jentsch, M.~Anand, and C.~T. Bauch, ``{Prioritising COVID-19 vaccination in
  changing social and epidemiological landscapes},'' \emph{medRxiv}, 2020.

\bibitem{meyers2005network}
L.~A. Meyers, B.~Pourbohloul, M.~E. Newman, D.~M. Skowronski, and R.~C.
  Brunham, ``{Network theory and SARS: predicting outbreak diversity},''
  \emph{Journal of theoretical biology}, vol. 232, no.~1, pp. 71--81, January
  2005.

\bibitem{hebert2020beyond}
L.~H{\'e}bert-Dufresne, B.~M. Althouse, S.~V. Scarpino, and A.~Allard,
  ``{Beyond R0: Heterogeneity in secondary infections and probabilistic
  epidemic forecasting},'' \emph{medRxiv}, 2020.

\bibitem{medlock2009optimizing}
J.~Medlock and A.~P. Galvani, ``{Optimizing influenza vaccine distribution},''
  \emph{Science}, vol. 325, no. 5948, pp. 1705--1708, September 2009.

\bibitem{prem2017projecting}
K.~Prem, A.~R. Cook, and M.~Jit, ``{Projecting social contact matrices in 152
  countries using contact surveys and demographic data},'' \emph{PLoS
  computational biology}, vol.~13, no.~9, p. e1005697, September 2017.

\bibitem{CDCvaccine}
{Centers for Disease Control And Prevention (CDC)}, ``How {CDC} is making
  {COVID-19} vaccine recommendations,''
  https://www.cdc.gov/coronavirus/2019-ncov/vaccines/recommendations-process.html.

\bibitem{vaccinechallenges}
``{WHO unveils global plan to fairly distribute COVID-19 vaccine, but
  challenges await},''
  https://www.sciencemag.org/news/2020/09/who-unveils-global-plan-fairly-distribute-covid-19-vaccine-challenges-await,
  {Science}.

\bibitem{britton2020mathematical}
T.~Britton, F.~Ball, and P.~Trapman, ``{A mathematical model reveals the
  influence of population heterogeneity on herd immunity to SARS-CoV-2},''
  \emph{Science}, vol. 369, no. 6505, pp. 846--849, August 2020.

\bibitem{tkachenko2020persistent}
A.~V. Tkachenko, S.~Maslov, A.~Elbanna, G.~N. Wong, Z.~J. Weiner, and
  N.~Goldenfeld, ``{Persistent heterogeneity not short-term overdispersion
  determines herd immunity to COVID-19},'' \emph{arXiv preprint
  arXiv:2008.08142}, 2020.

\bibitem{chang2020modelling}
S.~L. Chang, N.~Harding, C.~Zachreson, O.~M. Cliff, and M.~Prokopenko,
  ``Modelling transmission and control of the covid-19 pandemic in australia,''
  \emph{Nature communications}, vol.~11, no.~1, pp. 1--13, November 2020.

\bibitem{newman2002spread}
M.~E. Newman, ``{Spread of epidemic disease on networks},'' \emph{Physical
  review E}, vol.~66, no.~1, p. 016128, July 2002.

\bibitem{allard2020role}
A.~Allard, C.~Moore, S.~V. Scarpino, B.~M. Althouse, and
  L.~H{\'e}bert-Dufresne, ``{The role of directionality, heterogeneity and
  correlations in epidemic risk and spread},'' \emph{arXiv preprint
  arXiv:2005.11283}, 2020.

\bibitem{chen2020time}
Y.-C. Chen, P.-E. Lu, C.-S. Chang, and T.-H. Liu, ``{A Time-dependent SIR model
  for COVID-19 with undetectable infected persons},'' \emph{IEEE Transactions
  on Network Science and Engineering}, October-December 2020.

\bibitem{allard2009heterogeneous}
A.~Allard, P.-A. No{\"e}l, L.~J. Dub{\'e}, and B.~Pourbohloul, ``{Heterogeneous
  bond percolation on multitype networks with an application to epidemic
  dynamics},'' \emph{Physical Review E}, vol.~79, no.~3, p. 036113, March 2009.

\bibitem{allard2015general}
A.~Allard, L.~H{\'e}bert-Dufresne, J.-G. Young, and L.~J. Dub{\'e}, ``{General
  and exact approach to percolation on random graphs},'' \emph{Physical Review
  E}, vol.~92, no.~6, p. 062807, December 2015.

\bibitem{miller2012edge}
J.~C. Miller, A.~C. Slim, and E.~M. Volz, ``{Edge-based compartmental modelling
  for infectious disease spread},'' \emph{Journal of the Royal Society
  Interface}, vol.~9, no.~70, pp. 890--906, October 2011.

\bibitem{matrajt2020vaccine}
L.~Matrajt, J.~Eaton, T.~Leung, and E.~R. Brown, ``{Vaccine optimization for
  COVID-19, who to vaccinate first?}'' \emph{medRxiv}, 2020.

\bibitem{bubar2020model}
K.~M. Bubar, S.~M. Kissler, M.~Lipsitch, S.~Cobey, Y.~Grad, and D.~B.
  Larremore, ``{Model-informed COVID-19 vaccine prioritization strategies by
  age and serostatus},'' \emph{medRxiv}, 2020.

\bibitem{buckner2020optimal}
J.~H. Buckner, G.~Chowell, and M.~R. Springborn, ``{Optimal dynamic
  prioritization of scarce COVID-19 vaccines},'' \emph{medRxiv}, 2020.

\bibitem{li2020substantial}
R.~Li, S.~Pei, B.~Chen, Y.~Song, T.~Zhang, W.~Yang, and J.~Shaman,
  ``{Substantial undocumented infection facilitates the rapid dissemination of
  novel coronavirus (SARS-CoV-2)},'' \emph{Science}, vol. 368, no. 6490, pp.
  489--493, May 2020.

\bibitem{van2002reproduction}
P.~Van~den Driessche and J.~Watmough, ``{Reproduction numbers and sub-threshold
  endemic equilibria for compartmental models of disease transmission},''
  \emph{Mathematical biosciences}, vol. 180, no. 1-2, pp. 29--48,
  November-December 2002.

\bibitem{shao2009structure}
J.~Shao, S.~V. Buldyrev, L.~A. Braunstein, S.~Havlin, and H.~E. Stanley,
  ``{Structure of shells in complex networks},'' \emph{Physical Review E},
  vol.~80, no.~3, p. 036105, September 2009.

\bibitem{wang2014epidemic}
W.~Wang, M.~Tang, H.-F. Zhang, H.~Gao, Y.~Do, and Z.-H. Liu, ``{Epidemic
  spreading on complex networks with general degree and weight
  distributions},'' \emph{Physical Review E}, vol.~90, no.~4, p. 042803,
  October 2014.

\bibitem{mossong2008social}
J.~Mossong, N.~Hens, M.~Jit, P.~Beutels, K.~Auranen, R.~Mikolajczyk,
  M.~Massari, S.~Salmaso, G.~S. Tomba, J.~Wallinga \emph{et~al.}, ``{Social
  contacts and mixing patterns relevant to the spread of infectious
  diseases},'' \emph{PLoS Med}, vol.~5, no.~3, p. e74, March 2008.

\bibitem{davies2020age}
N.~G. Davies, P.~Klepac, Y.~Liu, K.~Prem, M.~Jit, and R.~M. Eggo,
  ``{Age-dependent effects in the transmission and control of COVID-19
  epidemics},'' \emph{Nature medicine}, vol.~26, no.~8, pp. 1205--1211, June
  2020.

\bibitem{wang2020clinical}
D.~Wang, B.~Hu, C.~Hu, F.~Zhu, X.~Liu, J.~Zhang, B.~Wang, H.~Xiang, Z.~Cheng,
  Y.~Xiong \emph{et~al.}, ``{Clinical characteristics of 138 hospitalized
  patients with 2019 novel coronavirus--infected pneumonia in Wuhan, China},''
  \emph{Jama}, vol. 323, no.~11, pp. 1061--1069, February 2020.

\bibitem{eikenberry2020mask}
S.~E. Eikenberry, M.~Mancuso, E.~Iboi, T.~Phan, K.~Eikenberry, Y.~Kuang,
  E.~Kostelich, and A.~B. Gumel, ``{To mask or not to mask: Modeling the
  potential for face mask use by the general public to curtail the COVID-19
  pandemic},'' \emph{Infectious Disease Modelling}, April 2020.

\bibitem{Moderna}
``{FDA Briefing Document Moderna COVID-19 Vaccine},''
  https://www.fda.gov/media/144434/download, December 2020.

\bibitem{polack2020safety}
F.~P. Polack, S.~J. Thomas, N.~Kitchin, J.~Absalon, A.~Gurtman, S.~Lockhart,
  J.~L. Perez, G.~P{\'e}rez~Marc, E.~D. Moreira, C.~Zerbini \emph{et~al.},
  ``{Safety and efficacy of the BNT162b2 mRNA Covid-19 vaccine},'' \emph{New
  England Journal of Medicine}, vol. 383, no.~27, pp. 2603--2615, December
  2020.

\end{thebibliography}
\end{document}